\documentclass[letterpaper,twocolumn,10pt]{article}
\usepackage{usenix2019_v3}

\usepackage{tikz}
\usepackage{amsmath}

\usepackage{enumitem}
\usepackage{subfig}
\usepackage{amsmath}
\usepackage{makecell}
\usepackage{caption}
\usepackage{graphicx}
\usepackage{float}
\usepackage{array}
\usepackage{booktabs}
\definecolor{azure_blue}{rgb}{0.0, 0.5, 1.0}
\definecolor{darkgreen}{rgb}{0.0, 0.7, 0}
\definecolor{darkred}{rgb}{0.6,0,0}

\usepackage{listings}
\definecolor{dkgreen}{rgb}{0,0.6,0}
\definecolor{gray}{rgb}{0.5,0.5,0.5}
\definecolor{mauve}{rgb}{0.58,0,0.82}
\definecolor{codeyellow}{rgb}{1.0, 0.49, 0.0}

\makeatletter 
\newcommand\notsotiny{\@setfontsize\notsotiny\@vipt\@viipt}
\makeatother
\usepackage{listings}
\definecolor{dkgreen}{rgb}{0,0.6,0}
\definecolor{gray}{rgb}{0.5,0.5,0.5}
\definecolor{mauve}{rgb}{0.58,0,0.82}
\definecolor{codeyellow}{rgb}{1.0, 0.49, 0.0}
\usepackage{tikz}
\newcommand*\circleBlack[1]{\tikz[baseline=(char.base)]{
            \node[shape=circle,fill,color=black,text=white,inner sep=0.05pt](char){#1};}}

\newcommand{\squishlist}{
  \begin{list}{$\bullet$}{
    \setlength{\itemsep}{0pt}       \setlength{\parsep}{3pt}
    \setlength{\topsep}{3pt}        \setlength{\partopsep}{0pt}
    \setlength{\leftmargin}{1em}    \setlength{\labelwidth}{1em}
    \setlength{\labelsep}{0.5em} } }

\newcommand{\squishend}{
  \end{list} }

\newcommand{\sysname}{{FlexiNS}}
\newcommand{\us}{{$\mu$s} }

\begin{document}
\begin{sloppypar}
    
\date{}

\title{\sysname: A SmartNIC-Centric, Line-Rate and Flexible Network Stack}

\author{
{\rm Xuzheng Chen$^{1}$} \and 
{\rm Jie Zhang$^{1}$} \and
{\rm Baolin Zhu$^{1}$} \and 
{\rm Xueying Zhu$^{1}$} \and
{\rm Zhongqing Chen$^{2}$} \and 
{\rm Shu Ma$^{2}$} \and
{\rm Lingjun Zhu$^{2}$} \and 
{\rm Chao Shi$^{2}$} \\ $^1$ Zhejiang University \and
{\rm Yin Zhang$^{1}$}\\ $^2$ Alibaba Cloud  \and 
{\rm Zeke Wang$^{1}$} 
}
\maketitle
\begin{abstract}
As the gap between network and CPU speeds rapidly increases, the CPU-centric network stack proves inadequate due to excessive CPU and memory overhead. While hardware-offloaded network stacks alleviate these issues, they suffer from limited flexibility in both control and data planes. Offloading network stack to off-path SmartNIC seems promising to provide high flexibility; however, throughput remains constrained by inherent SmartNIC architectural limitations.

To this end, we design \sysname{}, a SmartNIC-centric network stack with software transport programmability and line-rate packet processing capabilities. To grapple with the limitation of SmartNIC-induced challenges, \sysname{} introduces: (a) a header-only offloading TX path; (b) an unlimited-working-set in-cache processing RX path; (c) a high-performance DMA-only notification pipe; and (d) a programmable offloading engine. We prototype \sysname{} using Nvidia BlueField-3 SmartNIC and provide out-of-the-box RDMA IBV verbs compatibility to users. \sysname{} achieves 2.2$\times$ higher throughput than the microkernel-based baseline in block storage disaggregation and 1.3$\times$ higher throughput than the hardware-offloaded baseline in KVCache transfer.
\end{abstract}

\vspace{-2ex}
\section{Introduction}
\label{sec:intro}
\vspace{-2ex}

In the modern cloud, the demand for network speed grows quickly, with 200/400 Gigabit Ethernet (GbE) network interface controllers (NICs) widely deployed~\cite{cx6,cx7} and 800 GbE expected in the near future~\cite{cx8}. Consequently, the network stack must deliver elevated processing capacity. Moreover, as deployments vary broadly (e.g., single/multi-tenant, lossless/lossy fabric~\cite{irn, SRNIC, freeflow, masq}), and upper-layer applications expand from HPC and disaggregated storage~\cite{solar,os2g,azure-2} to GPU communication~\cite{alibaba-hpn, meta-rdma}, enhanced flexibility and programmability within the network stack become indispensable.

\textbf{CPU-centric network stacks}, represented by monolithic kernel-based~\cite{LITE, KRCore,qp_sharing_1} and microkernel-based~\cite{Snap, shenango, NetKernel, shinjuku, freeflow, tas, caladan, zygos,roud} designs, offer certain programmability via high-level software language (i.e., C/C++). However, Moore's law is slowing down, more portions of the CPU resources and memory bandwidth have to be occupied to handle increasing network bandwidth~\cite{towards_new_kernel,understand-host-network}. Besides, co‑locating the network stack with other CPU workloads induces cache and memory interference, resulting in high tail latency~\cite{shenango, caladan}.

\textbf{Hardware-offloaded network stacks}, represented by Remote Direct Memory Access (RDMA)~\cite{rdmaresources,rocev2} and TCP Offload Engine (TOE)~\cite{FlexTOE,acceltcp}, offer a compelling path to achieve high throughput in the post-Moore's Law era. Unfortunately, fixed‑function hardware cannot meet the various requirements of numerous different upper-layer applications~\cite{justitia, harmonic,masq,ovs, solar,alibaba-hpn} and rapid release cycles demanded by cloud environments~\cite{Snap, solar}. For example, RDMA NIC only supports limited transport protocols and is always treated as a black box, offering only limited ``configurable'' capabilities that are not truly programmable. Large cloud vendors collaborate with NIC vendors to integrate customized functions into next-gen NICs, which usually takes several years. 
FPGA-based NIC designs can offer a certain programmability using hardware description languages such as Verilog~\cite{1RMA, SRNIC,acceltcp, StaR, tonic}, but they still need a much longer development cycle and fail to meet the rapidly evolving demands of the cloud workloads~\cite{Snap, solar}. This leads to an interesting question: \textbf{How to build a line-rate, low-overhead network stack while guaranteeing software programmability to modern cloud providers?}

One promising solution is the SmartNIC-centric network stack, which allows offloading host workloads to the specially optimized SmartNIC processing units for better performance/cost efficiency. According to the location of the processing unit, SmartNIC is categorized into two types~\cite{ipipe}: on-path and off-path.
The former uses the dedicated processing framework and can only be programmed with low-level vendor-specific Microcode, posing significant burdens on developers~\cite{ipads_bf2, xenic}. In contrast, the latter features an on-NIC Arm SoC~\cite{bf2,bf3,broadcom_ps1100r} and Linux operating system, developers can seamlessly migrate existing programs. Therefore, we adopt the off-path SmartNIC design due to its flexibility. 

Although the off-path SmartNIC provides the modern cloud required programmability, building a line-rate network stack with off-path SmartNIC is non-trivial due to three identified unique challenges:

\noindent\textbf{1. High bandwidth contention on the Arm-NIC switch link. }
There is a NIC switch that connects the NIC, the Arm, and the host interface in the off-path SmartNIC. When outgoing traffic and incoming traffic both travel through the Arm, the link bandwidth between the Arm and the NIC switch easily becomes the bottleneck. If each NIC switch endpoint provides a 400 Gbps duplex bandwidth, a 400 Gbps outgoing traffic from host $\rightarrow$ Arm $\rightarrow$ NIC would fully saturate the link bandwidth of the Arm endpoint, and there is no link bandwidth for incoming traffic, which fails to achieve duplex line-rate. 

\noindent\textbf{2. Constrained Arm memory bandwidth. }
The network stack requires traffic to be temporarily staged in Arm memory, where both TX and RX paths incur two memcpy operations. Consequently, sustaining a 400 Gbps line rate demands about 1600 Gbps memory bandwidth\textemdash a requirement that far exceeds the capability of modern SmartNICs~\cite{bf2,bf3,broadcom_ps1100r, ipu}, which are commonly equipped with two DDR5 channels and provide about 500 Gbps memory bandwidth. Our empirical evaluation using ``Echo Server'' shows that this limitation can degrade the attainable throughput by 3.3$\times$.

\noindent\textbf{3. Long latency of small packet due to high PCIe interconnect latency. }
To invoke the network stack function, the Work Queue Entry (WQE) and Completion Queue Entry (CQE) are not directly passed to the NIC but detoured to the Arm, which introduces extra PCIe interconnect latency and is fatal for latency-sensitive applications. We test on the L2 Reflector application with 64B payload size, the Arm detour incurs an average 10.1\us latency, which is $2.2\times$ higher than the RDMA NIC and $1.4\times$ higher than the microkernel-based design.

In this paper, we design and implement \sysname{}, a SmartNIC-centric network stack with software transport programmability and line-rate packet processing capabilities. To address the above three challenges, \sysname{} introduces: 1) a header-only offloading TX path that constructs the custom packet header on Arm and integrates the host payload within NIC; 2) a unlimited-working-set in-cache processing RX path that receives packets into LLC, followed by cache self-invalidation after transferring the payload to the host destination; 3) a DMA-only notification pipe that leverages high-performance DMA engines to communicate between host CPU and Arm; and 4) a programmable offloading engine that empowers cloud providers to offload specialized tasks like customize one-sided operations and network functions. In sum, \sysname{} achieves line‑rate throughput and offers cloud providers the required software transport programmability.

We built \sysname{} on an off-the-shelf Nvidia Bluefield-3 SmartNIC with 400 GbE~\cite{bf3} connectivity and provided out-of-the-box RDMA IBV verbs compatibility, which is important for applying RDMA-aware optimizations. Benefitting from our novel design, \sysname{} can sustain full‑duplex line‑rate and comparable single-flow throughput with RDMA NIC. \sysname{} achieves 2.2$\times$ higher IOPS than the microkernel-based baseline in disaggregate block storage~\cite{luna} and 1.3$\times$ higher throughput than the hardware-offloaded baseline in KVCache transfer~\cite{mooncake}. 

\vspace{-2ex}
\section{Background and Motivation}
\label{sec:background}
\vspace{-2ex}

In modern data center, an ideal network stack should be capable of 1) providing high network throughput and low latency; 2) minimizing the host overhead for stack processing (i.e., CPU occupation); 3) reducing memory subsystem pressure and contention while other co-located applications are running simontaneously; 4) providing high programmability to allow developers to customize data plane policy (transport protocol and multi-path) and control plane policy (QoS, access control and congestion control). However, the existing designs fail to meet all the requirements as shown in Table~\ref{tab_comparision_networkstack}.

\begin{table}[]
	\centering
	\begin{scriptsize}
 	\caption{Comparison of network stack approaches.}
        \vspace{-2ex}
        \label{tab_comparision_networkstack}
	\begin{tabular}{|>{\centering\arraybackslash}m{1.8cm}|>{\centering\arraybackslash}m{1.1cm}|>{\centering\arraybackslash}m{1.1cm}|>{\centering\arraybackslash}m{1.2cm}|>{\centering\arraybackslash}m{1.2cm}|}
		\hline
		\textbf{Solution}   & \textbf{Throughput}  & \textbf{Host Overhead} 
           & \textbf{Memory Contention} & \textbf{\makecell{Programm-\\ability}}\\ 
		\hline
		\hline
          \textbf{Monolithic kernel-based~\cite{LITE}} & \textcolor{red}{low}  &  \textcolor{red}{High}  & \textcolor{red}{High} & \textcolor{orange}{Medium} \\
		\hline
		\textbf{Microkernel-based~\cite{Snap}} & \textcolor{orange}{Medium} &  \textcolor{orange}{Medium} & \textcolor{red}{High} & \textcolor{darkgreen}{High}\\
		\hline
		\textbf{Hardware-offloaded~\cite{cx7,1RMA}} & \textcolor{darkgreen}{High} &  \textcolor{darkgreen}{Low} & \textcolor{darkgreen}{None} & \textcolor{red}{Low}\\
		\hline
   		\textbf{Na\"ive \sysname} & \textcolor{red}{Low} & \textcolor{darkgreen}{Low} & \textcolor{darkgreen}{None} & \textcolor{darkgreen}{High}\\
		\hline        
   		\textbf{\sysname} & \textcolor{darkgreen}{High} & \textcolor{darkgreen}{Low} & \textcolor{darkgreen}{None} & \textcolor{darkgreen}{High} \\
		\hline    
	\end{tabular}
	\end{scriptsize}
    \vspace{-2ex}
\end{table}

\begin{figure}
    \centering
    {\includegraphics[keepaspectratio=true, width=0.98\linewidth]{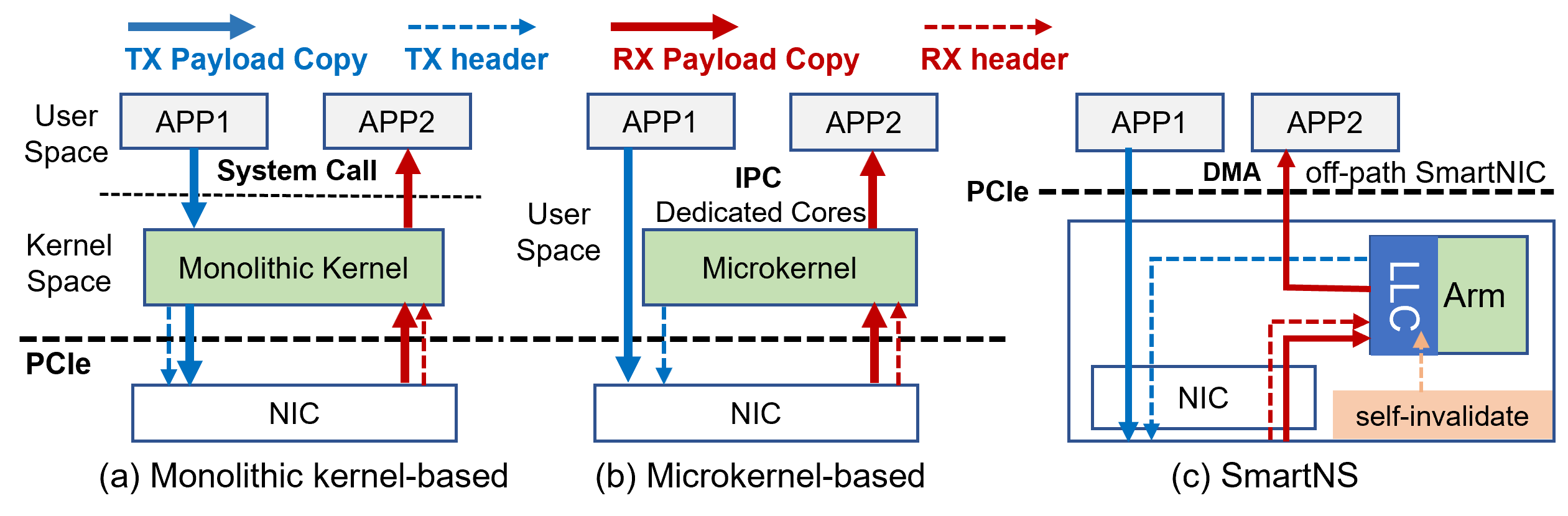}
    \vspace{-1ex}
    \caption{Comparison of different network stack designs.}
    \label{fig_overall_view}}
    \vspace{-4ex}
\end{figure}

\vspace{-2ex}
\subsection{CPU-centric Network Stack}
\label{sec:cpu-centric-network}
\vspace{-1ex}
\subsubsection{Monolithic Kernel-based Network Stack}
\vspace{-1ex}
Figure~\ref{fig_overall_view}a illustrates the data flow of a monolithic kernel-based network stack. In the TX path, the user application invokes a system call to transition to the kernel mode, where the network stack processes protocols and constructs packet headers. Then, it copies the user's payload into a pinned DMA-safe memory region and notifies the NIC to transmit. In the RX path, the NIC places received packets in a kernel-space reception queue and triggers an interrupt. After protocol processing, the network stack copies the payload to the user buffer and signals the user application. The system call overhead can incur a slowdown of up to 75\%, as reported in prior research~\cite{KRCore, LITE, NetKernel}. Besides, the high complexity of developing kernel code significantly harms the programmability of monolithic kernel-based network stacks.

\vspace{-3ex}
\subsubsection{Microkernel-based Network Stack}
\vspace{-1ex}
To mitigate the aforementioned host CPU overhead and programming complexity, many systems adopt the microkernel-based network stack~\cite{Snap, shenango, NetKernel, shinjuku, freeflow, tas, caladan, zygos} to replace costly system calls with lightweight IPC (Inter-Process Communication), as illustrated in Figure~\ref{fig_overall_view}b. There are three main benefits. First, compared with system calls, the overhead of IPC in modern CPUs is much lower as it preserves the application cache locality~\cite{Snap}. Second, it allows applications to directly use a pinned DMA-safe memory region, avoiding the memcpy overhead in the TX path. Third, a microkernel-based network stack has much higher programmability, as developing and deploying userspace code is much easier.

\vspace{-3ex}
\subsubsection{Issues of CPU-centric Network Stack}
\vspace{-1ex}
As Moore's law is slowing down, the gap between network and CPU speeds is rapidly increasing. With 200/400 GbE NICs already deployed in modern datacenters~\cite {AWSSRD,azure-2,alibaba-hpn} and 800 GbE expected in the near future~\cite{cx8}, both CPU-centric designs face two key issues.

\noindent\textbf{I1: High host CPU overhead. }
The host CPU overhead mainly consists of payload copy, which spends more than 60\% of CPU resources~\cite{understand-host-network,towards_new_kernel}. Due to the limited Line Fill Buffer~(LFB) per CPU core~\cite{understanding_host_network}, it can achieve only about 7GB/s memcpy speed per core. Cai et al.~\cite{towards_new_kernel} propose the decoupling data copy from application threads and using multi-threaded memcpy, but it still needs more than 15 cores to achieve the 400Gbps memcpy without protocol processing.

To illustrate the host CPU overhead, we construct an ``Echo Server" using a monolithic kernel-based~\cite{KRCore} and a microkernel-based~\cite{Snap} network stack, respectively, and compare with the RDMA NIC~\cite{fasst}. This echo server has 8 memory DDR5 channels and achieves around 160GB/s memory bandwidth. Both machines are equipped with a 400 GbE NVIDIA ConnectX-7 NIC~\cite{cx7}, and the packet size is 2KB, each CPU core serves 8 connections with 64 TX depth.

\begin{figure}[]
    \subfloat[Network throughput]{
        \label{fig_e_motivation_core_throughput}
        \includegraphics[keepaspectratio=true, width=0.48\linewidth]{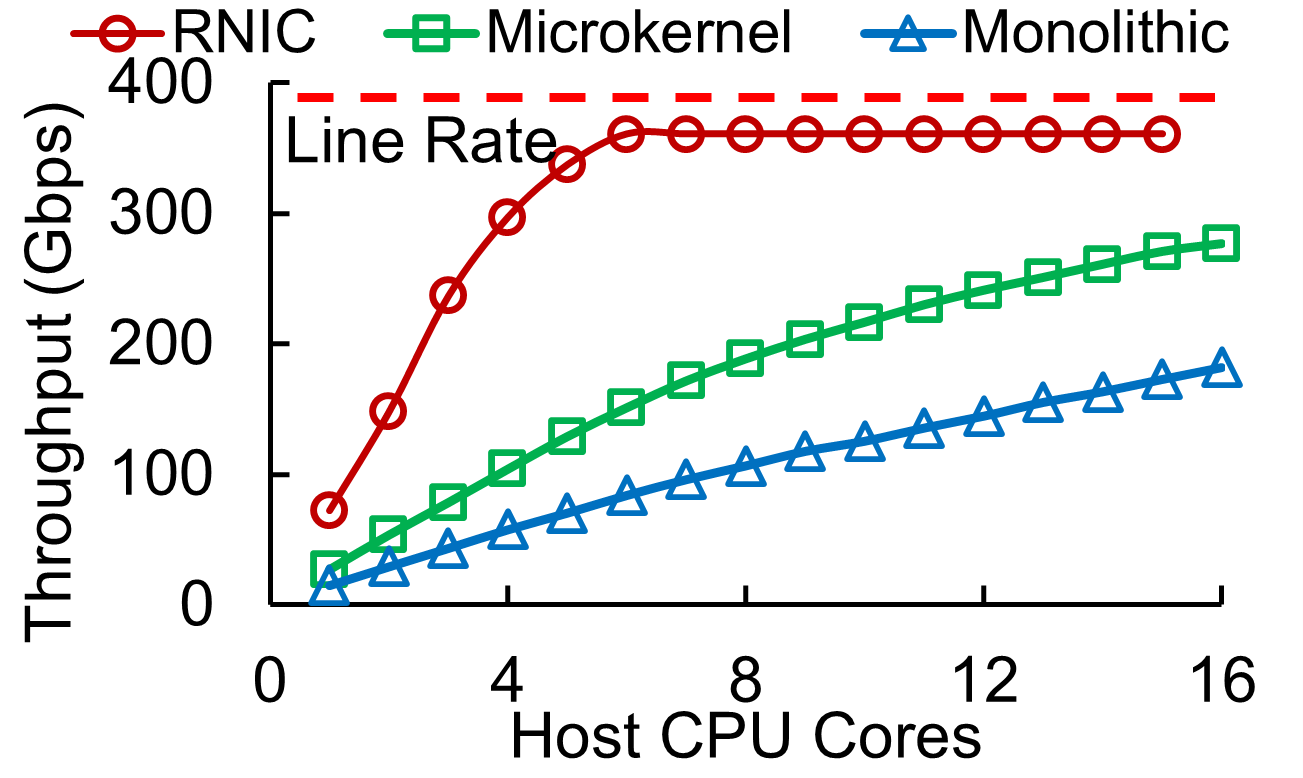}
    }
    \subfloat[Host memory bandwidth]{
        \label{fig_e_motivation_throughput_bw}
        \includegraphics[keepaspectratio=true, width=0.48\linewidth]{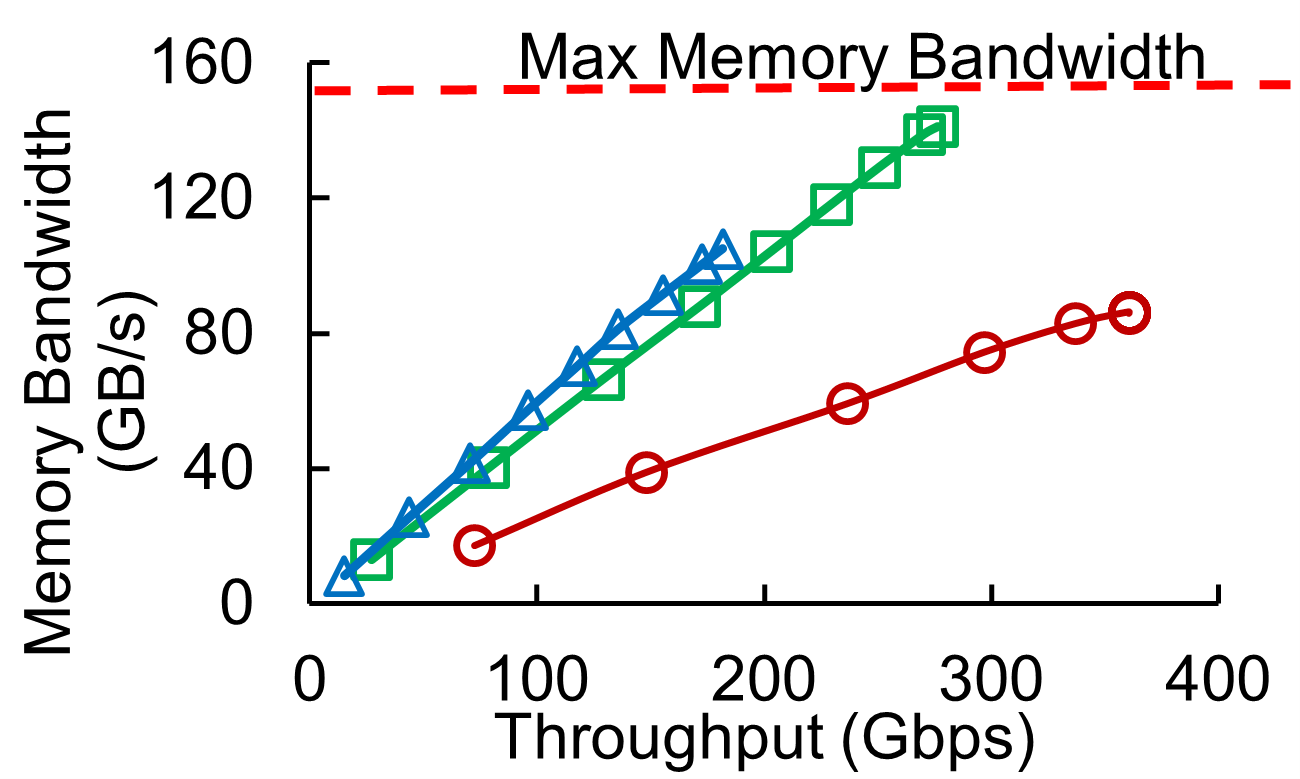}
    }
    \vspace{-1ex}
    \caption{Comparison of different stacks achieved throughput and corresponding host memory bandwidth usage.}
    \label{fig_e_motivation_core_bw}
    \vspace{-4ex}
\end{figure}

Figure~\ref{fig_e_motivation_core_throughput} illustrates the throughput under different host CPU cores. We observe that due to expensive system calls, the monolithic kernel exhibits approximately 42\% lower throughput than the microkernel-based design, which achieves only around 39\% per-CPU-core throughput compared with RDMA NIC, mainly due to the additional memcpy in RX path. We predict that this performance disparity would be widened as network speeds approach 800 Gbps in the near future.





\noindent\textbf{I2: High host memory contention. } 
The growth of network speed also places contention on the memory subsystem, which has been extensively explored in prior works~\cite{smartds, idio, resq}.

On the one hand, the extra memcpy adds significant memory pressure to the network stack. For example, only one extra memcpy in the RX path of the 400Gbps network needs at least 200GB/s memory bandwidth to achieve full-duplex line-rate throughput, which requires at least 8 DDR5-4800 memory channels (typical single‑socket Intel servers support up to 8 channels~\cite{demystifying_cxl,ddr5}). As shown in Figure~\ref{fig_e_motivation_throughput_bw}, both monolithic kernel and microkernel designs consume roughly 1.9$\times$ memory bandwidth compared with the RDMA NIC. The monolithic kernel almost fully saturates the available memory bandwidth when its throughput reaches 300Gbps. 

\begin{figure}[]
    \subfloat[Monolithic kernel-based system]{
        \label{fig_e_motivation_interference_monolithic}
        \includegraphics[keepaspectratio=true, width=0.48\linewidth]{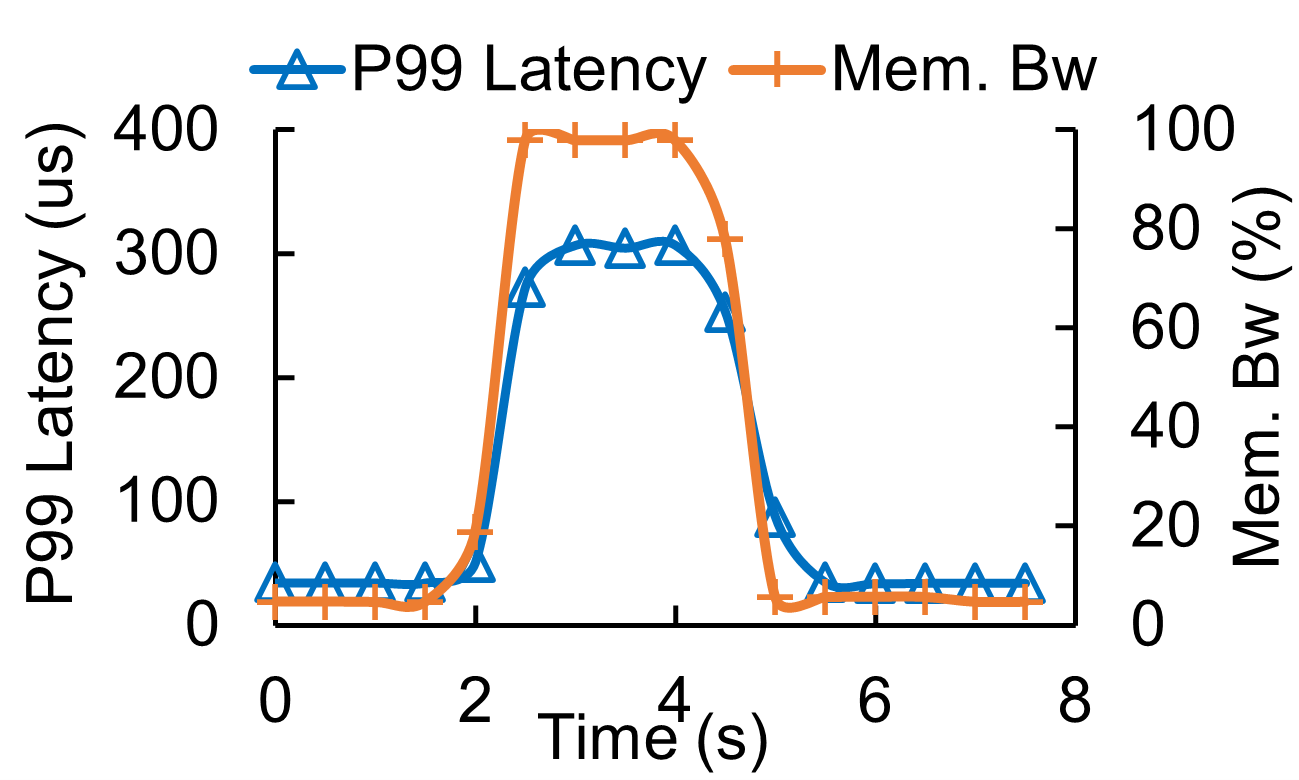}
    }
    \subfloat[Microkernel-based system]{
        \label{fig_e_motivation_interference_micro}
        \includegraphics[keepaspectratio=true, width=0.48\linewidth]{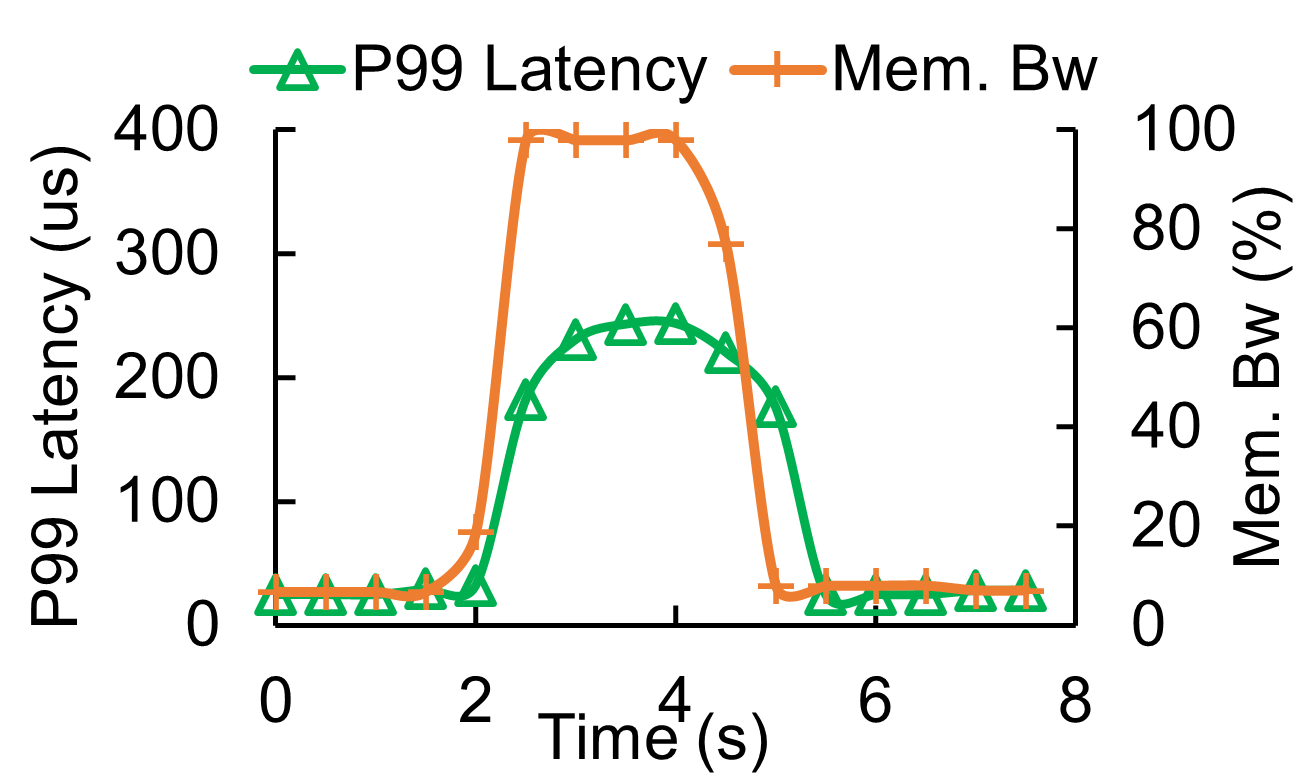}
    }
    \vspace{-1ex}    
    \caption{Memory-intensive application causes interference with the network stack.}
    \label{fig_e_motivation_interference}
    \vspace{-4ex}
\end{figure}

On the other hand, CPU‑centric network stacks suffer severe interference from co‑located memory‑intensive applications in multi‑tenant environments. As the CPU memory controller is shared among all cores, a core would suffer high memory access latency when other cores issue lots of memory transactions, resulting in high tail latency~\cite{caladan}. To quantify this effect, we use one CPU core as an echo server and other cores running Memory Latency Check~(MLC)~\cite{mlc} to simulate a memory-intensive garbage collection application. Figure~\ref{fig_e_motivation_interference} illustrates the P99 latency of the echo server, MLC starts at $2s$ and ends at $5s$. We observe that the interference increases P99 latency by $10.5\times$/$9.2\times$ for monolithic kernel and microkernel-based designs, respectively. This indicates that CPU-centric network stacks suffer from high tail latency in the case of co-located memory-intensive applications.

\vspace{-2ex}
\subsection{Hardware-offloaded Network Stack}
\vspace{-1ex}

To address the three aforementioned issues, hardware offloading technologies, e.g., RDMA, are used to offload the entire network stack to the NIC hardware to achieve high throughput, low latency, and low host overhead~\cite{cx7, FlexTOE, 1RMA, AWSSRD}. However, a hardwired network stack can not meet the various requirements of numerous different upper-layer applications (e.g., GPU communication, disaggregated storage, and high-performance computing). Specifically, their rigid architecture introduces programmability limitations in two dimensions. 

\noindent\textbf{I3: Inadequate data/control plane programmability. }
Commercial NICs only support a limited number of transports. For example, Nvidia ConnectX-7~\cite{cx7} only supports RoCEv2~\cite{rocev2} and InfiniBand~\cite{ib}. However, extensive innovations have been proposed to customize the transport layer to address specialized application requirements, like Google Snap~\cite{Snap} 1RMA~\cite{1RMA} and Falcon~\cite{Falcon}, AWS SRD~\cite{AWSSRD}, Alibaba Solar~\cite{solar, luna}, Tesla TTPoE~\cite{ttpoe}, and IRN~\cite{irn}. Likewise, modern data centers impose diverse control plane requirements: GPU communication needs high-precision congestion control~\cite{dcqcn, swift, hpcc, RevisitCC} and multi-path routing~\cite{AWSSRD,mprdma}; VPC needs multi-tenancy performance isolation~\cite{justitia,harmonic,masq,rdmaresources} and access control~\cite{ovs}. However, commercial RDMA NICs are always treated as black boxes and offer only limited ``configurable" capabilities that are not truly programmable. It takes years to collaborate with RNIC vendors to integrate these custom methods into the next-gen ASIC NIC.

To meet the need for programmability, researchers propose several FPGA-based NIC designs~\cite{SRNIC,1RMA, StaR, FlexTOE}, with which users can update the on-NIC transport module using hardware description languages~(HDL) such as Verilog. However, developing using HDL has a much longer development cycle, and fails to meet the rapidly evolving demands of the cloud workloads, which could update weekly or monthly~\cite{Snap, solar}. 

\begin{figure}
	\centering
	{\includegraphics[keepaspectratio=true, width=0.9\linewidth]{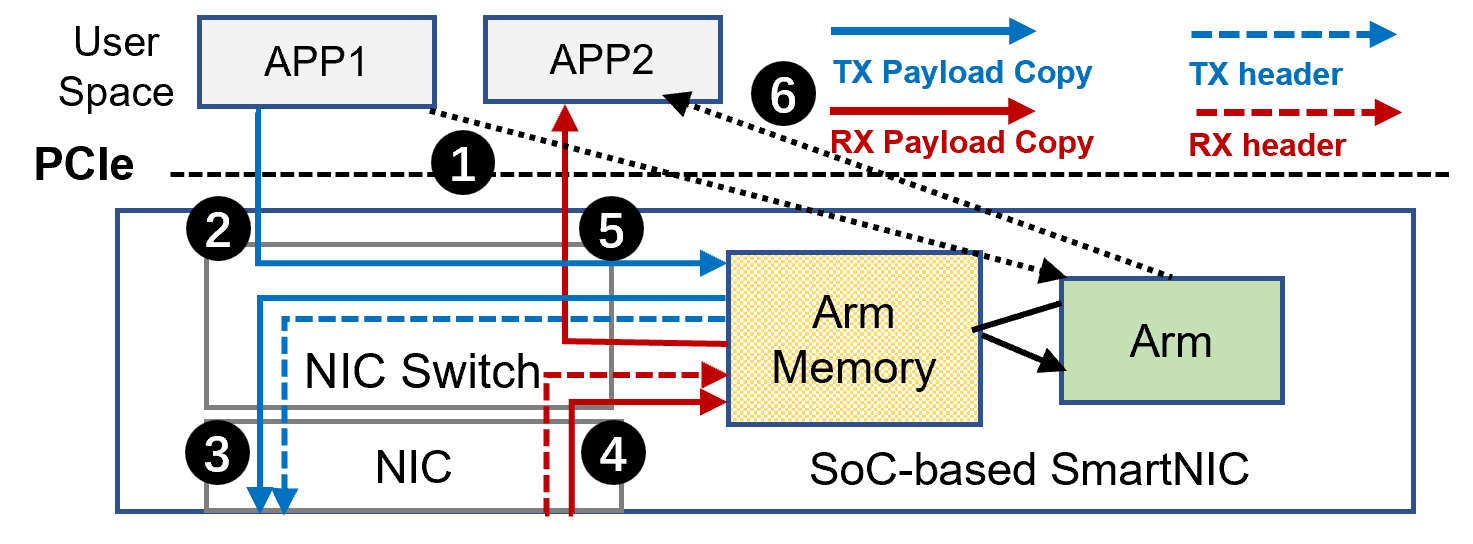}
        \vspace{-1ex}    
    \caption{High-level system architecture and TX/RX data flow of a na\"ive SmartNIC-centric network stack.}
	\label{fig_naive_smartns}}
    \vspace{-4ex}
\end{figure}

\vspace{-2ex}
\subsection{Na\"ive SmartNIC-centric Network Stack}
\label{sec:smartns-motivation}
\vspace{-1ex}

Modern off-path SmartNICs(e.g., NVIDIA BlueField~\cite{bf2,bf3}, Broadcom Stingray~\cite{broadcom_ps1100r}, Intel IPU~\cite{ipu}) feature an on-NIC programmable SoC (usually consisting of Arm), and thus provide an opportunity to offload the network stack to the NIC while preserving software programmability. 

Figure~\ref{fig_naive_smartns} sketches such a na\"ive SmartNIC-centric network stack design. 
In the TX path, the user application notifies Arm to send the packet (\circleBlack{1}), then Arm fetches the payload into its memory (\circleBlack{2}), prepares the packet header, and encapsulates the header and payload into the corresponding packet and send it out (\circleBlack{3}). 
In the RX path, the incoming packet is first stored in the Arm memory (\circleBlack{4}). Then the Arm parses the packet header, processes the protocol, and delivers the payload to the corresponding host buffer (\circleBlack{5}), at last the Arm signals the host CPU for notification (\circleBlack{6}). 
This na\"ive design allows implementing the entire transport layer in the Arm, thus providing software programmability while minimizing host CPU consumption and host memory bandwidth. However, it introduces three challenges in achieving high performance.

\noindent\textbf{C1: High bandwidth contention on the Arm-NIC switch link. }
An off-path SmartNIC features a NIC switch to connect the NIC, the Arm, and the host interface, as shown in Figure~\ref{fig_naive_smartns}. The Arm sits outside the host/NIC path, and that is why it is called off-path SmartNIC. In an off-path architecture, the link bandwidth between the Arm and NIC switch becomes the bottleneck when outgoing traffic and incoming traffic both travel through the Arm. If each NIC switch endpoint provides a 400 Gbps duplex bandwidth, a 400 Gbps outgoing traffic from host $\rightarrow$ Arm $\rightarrow$ NIC would fully saturate the link bandwidth of the Arm endpoint, and there would be no link bandwidth for incoming traffic.


\noindent\textbf{C2: Constrained Arm memory bandwidth. }
Compared to the host CPU, off-path SmartNICs offer comparable general-purpose computing capabilities but feature a significantly weaker memory subsystem~\cite{benchbf}. For instance, the BF3 SmartNIC features two 5600MT/s DDR5 channels to deliver the theoretical 716.8Gbps memory bandwidth, and its achievable mixed read-write bandwidth is approximately 480Gbps~\cite{benchbf}. However, the network stack requires network traffic to be temporarily staged in Arm DRAM, resulting in the payload going through device memory four times, as illustrated in Figure~\ref{fig_naive_smartns}. Thus, 400Gbps network throughput requires about 1600Gbps memory bandwidth\textemdash a requirement that far exceeds the capability of BF3. BlueField-2, Intel IPU, and Broadcom PS1100R also have similar memory bandwidth shortages. 

To illustrate this, we implement a na\"ive SmartNIC-centric network stack in Figure~\ref{fig_naive_smartns} and conduct the "Echo Server" experiments again (\S\ref{sec:cpu-centric-network}). We find that it achieves only 120 Gbps throughput (30\% of the network link throughput). Through the memory monitor on the Arm, we identify that the primary culprit behind its performance degradation is memory bandwidth exhaustion. 

\noindent\textbf{C3: Long latency of small packet due to high PCIe interconnect latency. } 
As shown in Figure~\ref{fig_naive_smartns}, to invoke the network stack function, the WQE and CQE are not directly passed to the NIC but detoured to the Arm, which introduces extra PCIe interconnect latency~(taking sub-microseconds). Although throughput-sensitive applications are unaffected by the additional latency, it still leads to performance degradation for latency-sensitive applications, particularly when the payload size falls within one MTU. We test on L2 Reflector application and use 64B payload size, the na\"ive SmartNIC-centric network stack demonstrated an unexpectedly high latency of roughly 10.1\us, which is $2.2\times$ higher than the RDMA NIC and $1.4\times$ higher than the microkernel-based network stack.



\vspace{-2ex}
\section{Design}
\vspace{-2ex}

\subsection{Overview}
\begin{figure}
	\centering
	{\includegraphics[keepaspectratio=true, width=0.95\linewidth]{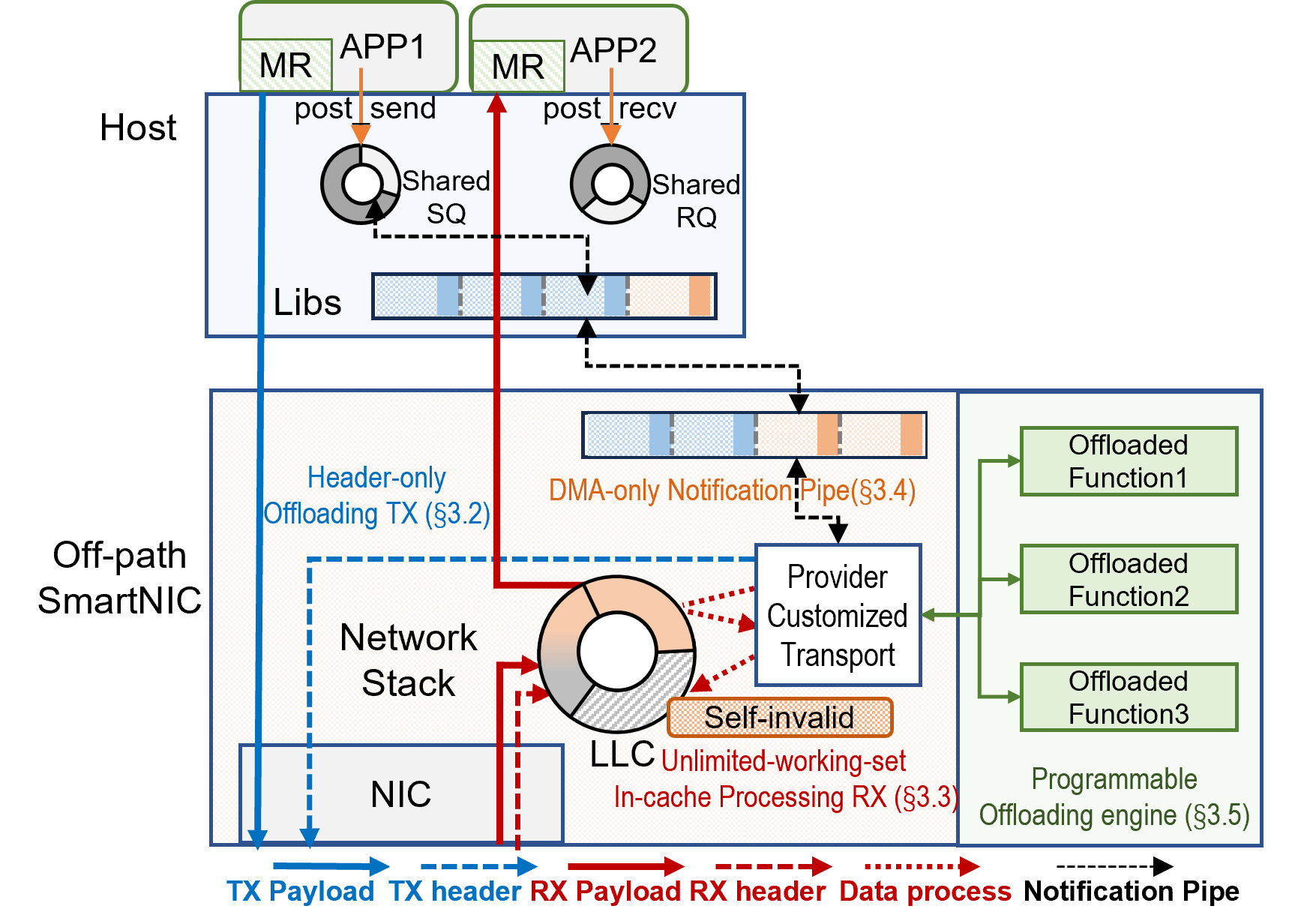}
        \vspace{-1ex}
        \caption{Architecture overview of \sysname{}.}
	\label{fig_smartns}}
    \vspace{-4ex}
\end{figure}

\vspace{-1ex}
To address the above three challenges, we propose \sysname{}, a SmartNIC-centric network stack with software transport programmability and line-rate packet processing capabilities. Figure~\ref{fig_smartns} presents the architecture overview of \sysname{}. The main network stack running on the on-NIC Arm cores as a separate user process consists of 1) a header-only offloading TX path (\S~\ref{sec:tx_path}) that constructs custom packet headers on the Arm and directly integrates the host payload within the NIC; 2) a unlimited-working-set in-cache processing RX path (\S~\ref{sec:rx_path}) that receives packets into the LLC, followed by cache self-invalidation after transferring the payload to the host destination; 3) a DMA-only notification pipe (\S~\ref{sec::dma_pipe}) that solely leverages high performance on-NIC DMA engines to notify the host or Arm core; and 4) a programmable offloading engine (\S~\ref{sec:one_side}) that allows cloud providers to offload specialized functions like customized one-sided operations and network functions. The Arm cores are divided into two partitions: the former executes flexible transport protocols processing, customizable congestion control algorithms (CCA), and QoS policies. Cloud providers are free to implement their customized transport protocols and congestion control algorithms in the Arm cores with the high-level software programming language (i.e., C/C++). In this paper, we have implemented two example protocols: RoCEv2~\cite{rocev2} and Solar~\cite{solar}, along with the widely adopted DCQCN~\cite{dcqcn} as the CCA. The latter executes cloud providers' offloading functions to mitigate the host CPU occupation. 

On the host side, \sysname{} provides 1) a user-space runtime library linked to each application; and 2) a kernel module that registers the entire system as an RDMA device and communicates with the network stack. This user library forwards control verbs~(like \texttt{create\_qp} and \texttt{modify\_qp}) to the kernel module and directly passes data verbs~(like \texttt{post\_send} and \texttt{post\_recv}) to Arm through DMA-only notification pipe. Since all operations are offloaded to the SmartNIC, this library introduces negligible overhead and interference for the host CPU. By registering the entire system as an RDMA device in kernel module, users can enjoy out-of-the-box RDMA IBV verbs compatibility~\cite{rdma-core}, which is important for applying RDMA-aware optimizations~\cite{hybird-rdma-wei,one-side-disaggregated, no-compromise, KRCore}.



\vspace{-2ex}
\subsection{Header-only Offloading TX Path}
\label{sec:tx_path}
\vspace{-1ex}

To address the challenge \#1 (\textbf{C1}), we propose the Header-only offloading TX path. First, we discuss why the following general and widely used TX path design is inadequate for achieving high throughput.


\begin{figure}
	\centering
	{\includegraphics[keepaspectratio=true, width=0.9\linewidth]{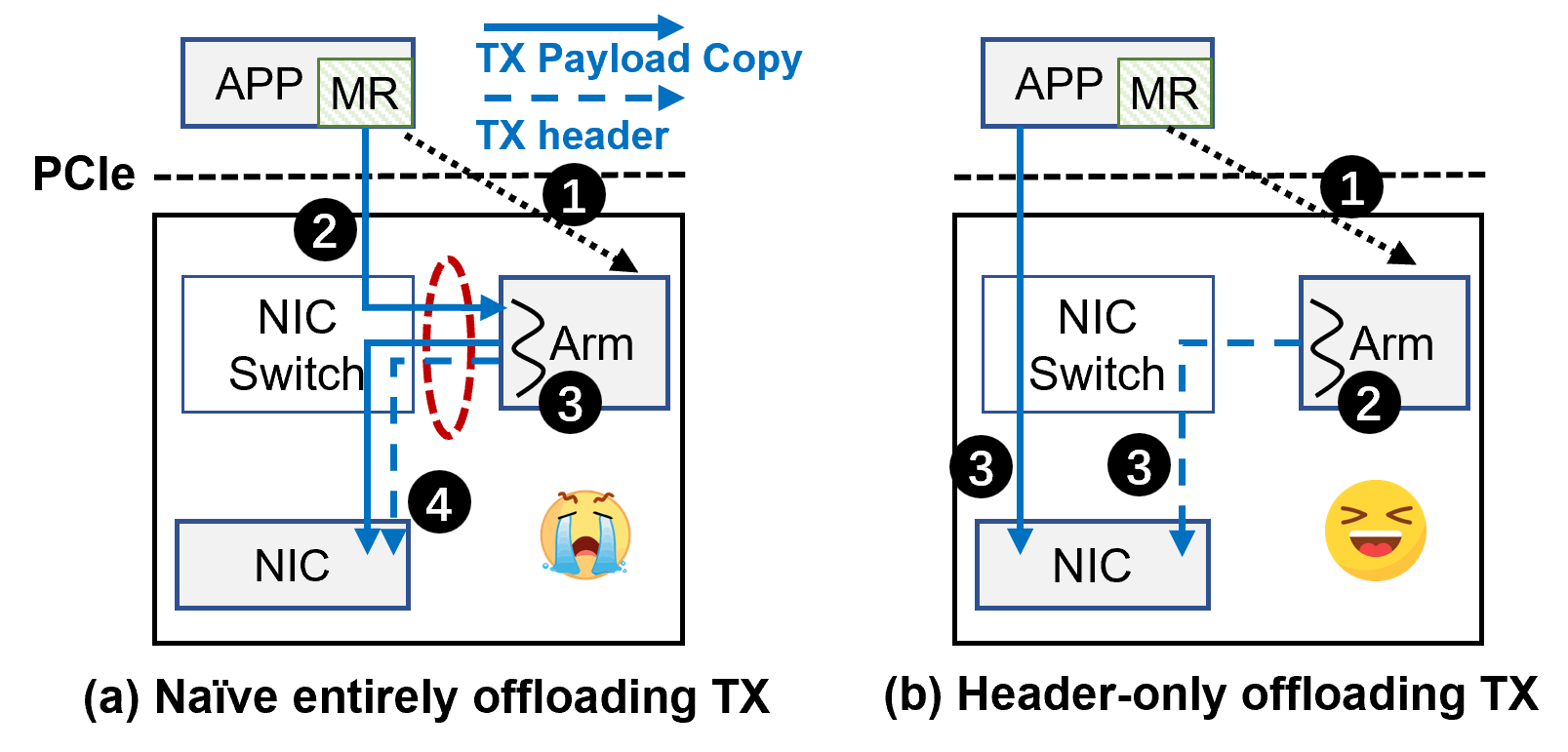}
        \vspace{-1ex}    
	\caption{Comparison of TX path strategies.}
	\label{fig_decoupled_tx_compare}}
    \vspace{-4ex}
\end{figure}



\noindent\textbf{Option: Na\"ive entirely offloading TX path. }
Figure~\ref{fig_decoupled_tx_compare}a depicts the process of na\"ive entirely offloading TX strategy, which has been widely adopted in prior studies~\cite{splitrpc, tcp_offload_bf2,os2g}. When the user application initiates packet transmission (\circleBlack{1}), the Arm utilizes intra-node DMA/RDMA~\cite{ipads_bf2,benchbf} to read the payload into the Arm memory (\circleBlack{2}). The Arm subsequently constructs the corresponding packet header and merges it with the payload into a contiguous buffer (\circleBlack{3}). At last, the Arm transmits the assembled packet to the network (\circleBlack{4}). This approach benefits from standard IBV verbs, making it straightforward to implement. However, this approach causes significant Arm memory subsystem pressure and saturates the Arm endpoint’s full‑duplex link bandwidth: although it achieves line rate for TX traffic, there would be no link bandwidth for RX traffic~(discussed in $\S$\ref{sec:smartns-motivation}).

\noindent\textbf{Our approach: Header-only offloading TX path. }
To address the limitations of the above design, we decouple the packet header and payload across from Arm memory and host memory, aiming to reduce the Arm endpoint's link occupation and minimize Arm memory bandwidth pressure. Our key insight is that the network stack typically neither accesses nor modifies the packet payload, thus, we can construct custom header on Arm and allow the payload be fetched directly by the NIC without any additional data movement, while leveraging NIC hardware for residual operations such as CRC computation, IP Checksum, and AES encryptio. Figure~\ref{fig_decoupled_tx_compare}b illustrates the process. After the Arm prepares the packet header (\circleBlack{2}), both the header and the payload are directly retrieved by the NIC (\circleBlack{3}). This simple yet effective design offers three key advantages: (1) eliminates the Arm-NIC switch link congestion, (2) reduces bandwidth pressure on the Arm memory subsystem, and (3) decreases the programming complexity of the network stack.

\begin{figure}
	\centering
	{\includegraphics[keepaspectratio=true, width=0.95\linewidth]{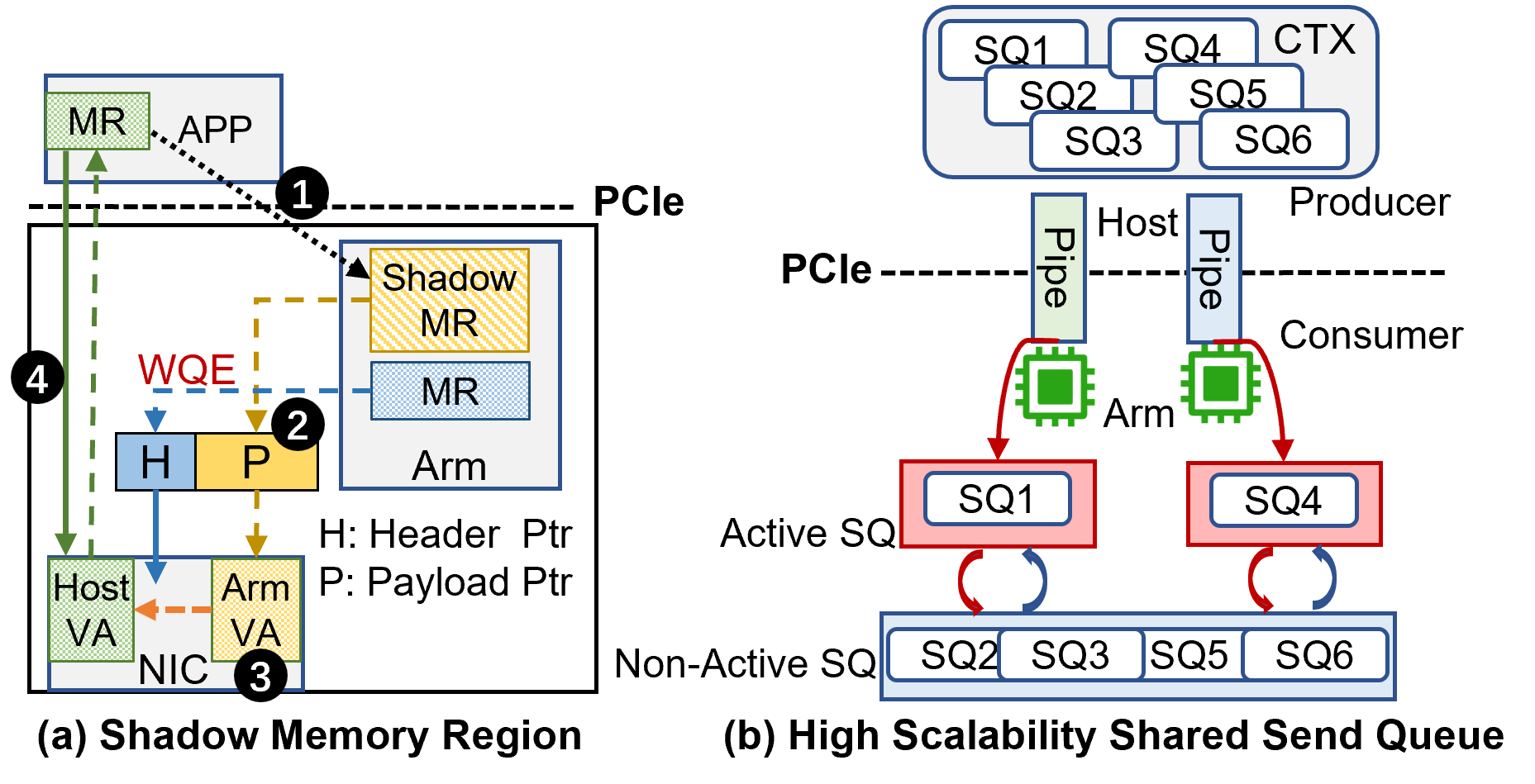}
        \vspace{-1ex}
	\caption{Architecture of Shadow Memory Region and High Scalability Shared Send Queue.}
	\label{fig_shadow_mr_shared_sq}}
        \vspace{-3ex}
\end{figure}

RDMA operations require the payload buffer to reside in a pre-registered memory region, each memory region associated with a specific RDMA context. Since the RDMA context includes hardware resources and can't be shared between Arm and host CPU, Arm can't directly use payload information pertaining to the host’s memory region to construct WQE. To expose host payload buffer information to the network stack, we introduce the \textbf{shadow memory region}. Each time the user application registers a memory region, the kernel module informs the (host VA, Size) to the Arm. The Arm then selects an unused virtual address range (Arm VA), instructs the NIC to establish a hardware mapping between the host VA and the Arm VA, designates the (Arm VA, Size) as the shadow memory region, and finally returns the Arm VA to the user library. Importantly, this Arm VA is not mapped to any physical address and cannot be directly accessed via load/store instructions, thereby introducing no actual memory overhead. As shown in Figure~\ref{fig_shadow_mr_shared_sq}a, when the user initiates packet transmission, the library automatically translates the host VA to Arm VA (\circleBlack{1}). Subsequently, the Arm utilizes the Arm VA in the shadow memory region to construct the WQE and transmit the packet (\circleBlack{2}), thus, NIC hardware resolves the corresponding host VA through that pre-established mapping (\circleBlack{3}), fetches the payload directly from host memory, and integrate the custom packet headers generated on Arm (\circleBlack{4}). This header-only offloading TX path ensures zero-copy while remaining transparent to the user application.

Another critical concern is the scalability of our software-based TX mechanism. Assuming each SQ is assigned a dedicated notification pipe, the overhead from software looping is negligible when the number of SQs is relatively small ($\leq$ 100). However, as the number of SQs increases, this approach becomes less efficient. The looping time for 2000 SQs can be up to 50 µs~\cite{qp_sharing_1}. To achieve low overhead and high scalability, we propose the \textbf{high-scalability shared send queue}. As shown in Figure~\ref{fig_shadow_mr_shared_sq}b, instead of allocating one pipe per SQ, we assign $N$ pipes per RDMA context, where $N$ equals the number of Arm cores. Each Arm core is responsible for polling a dedicated pipe within its assigned context. When a new SQ is created in a given context, the driver selects an underloaded Arm core and utilizes its corresponding pipe for SQE transmission. Given that RDMA-based systems typically create a limited number of contexts~\cite{hybird-rdma-wei, one-side-disaggregated, scalable_mpi}, each Arm core is required to poll only a small number of pipes. This sharing mechanism doesn't break the performance isolation of RDMA resources because RDMA hardware resources are allocated at the context level, and we only share notification pipes within the same RDMA context. Furthermore, to efficiently manage unfinished requests, we maintain an active SQ table that tracks unfinished operations for each Arm core. An SQ is added to the active SQ table when the Arm core receives an SQE related to that SQ, and this SQ will be transitioned to the non-active SQ pool when all the requests are finished.

\vspace{-3ex}
\subsection{Unlimited-working-set In-Cache \\Processing RX Path}
\label{sec:rx_path}
\vspace{-1ex}


\begin{figure*}
	\centering
	{\includegraphics[keepaspectratio=true, width=0.88\linewidth]{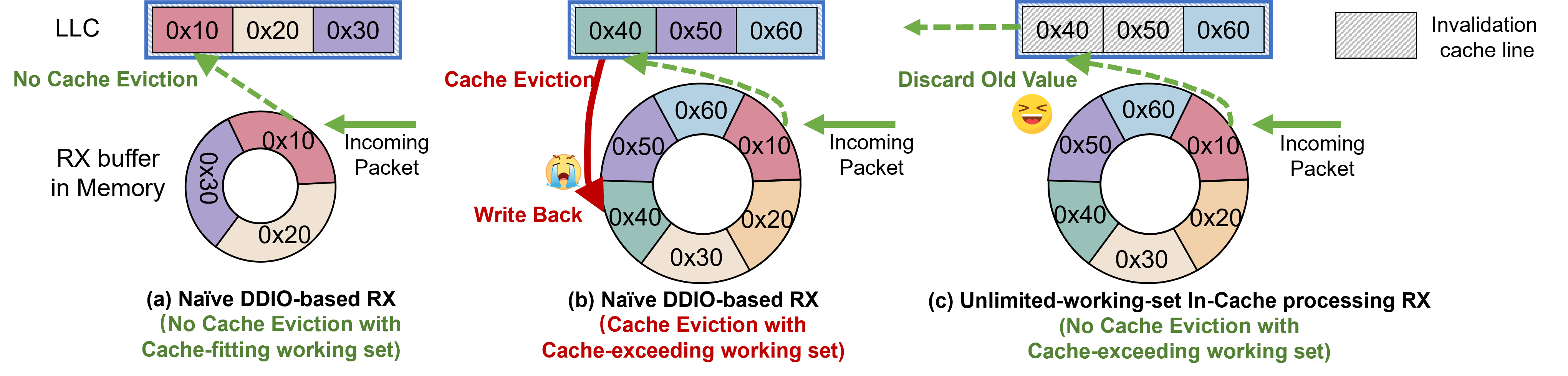}
        \vspace{-2ex}
	\caption{Comparison of RX path strategies.}
	\label{fig_cache_invalid}}
        \vspace{-3ex}
\end{figure*}

To address challenge \#2 (\textbf{C2}), we propose unlimited-working-set in-cache processing RX path. In the following, we discuss the challenges, followed by its design. 

\noindent\textbf{Challenges. }
Although TX path optimizations can greatly reduce the precious Arm memory bandwidth incurred by the TX traffic because TX payload does not traverse through the Arm memory, the limited Arm memory bandwidth is still insufficient to support line-rate RX traffic if all RX traffic passes through the Arm memory. Take BF3 as an example, we observe that the Arm memory bandwidth is exhausted when the RX traffic bandwidth is only 60\% of the network line rate in the previous "Echo Server" experiment. 


Intel DDIO is widely used in network applications to relieve memory bandwidth problems~\cite{ddio-2,ddio-3,ddio-4,benchmark-ddio,rdca}. DDIO allows PCIe devices such as NICs to write incoming packets to the LLC directly, thus avoiding memory bandwidth consumption. The Arm on the SmartNIC also supports the DDIO mechanism and can use the whole LLC space to serve the DDIO mechanism, thus, it can potentially address the above-mentioned memory bandwidth challenge. 
However, we find that the benefits of DDIO are highly constrained: \textbf{the working set size must be smaller than the cache size.}

\noindent\textbf{Cache-fitting working set. }
Figure~\ref{fig_cache_invalid}a demonstrates how a network stack receives packets when the working set (i.e., receive buffer) can fit in the cache. We observe that every address in the working set is cached, and an incoming packet can be directly written into the cache without cache eviction. In this case, the DDIO mechanism can well address the limited Arm memory bandwidth challenge.

\noindent\textbf{Cache-exceeding working set. }
However, in a network stack RX path, the working set (i.e., receive buffer) size is usually very large, which far exceeds the cache size. Ideally, the working set size only needs to be larger than the product of the network bandwidth and the average packet processing time. However, packet processing time varies significantly, and the packets may arrive in a burst manner (i.e., a sudden spike in packet arrivals)~\cite{benchmark-ddio,idio}. To avoid dropping packets, the network stack RX path must ensure that the working set size can absorb the burst. Consequently, multi-hundred-gigabit network stacks typically employ more than 512 descriptors per RX queue~\cite{benchmark-ddio, hybird-rdma-wei, shring}. Assuming that there is one RX queue per Arm core and the maximum packet size is 4 KB, the total working set size would be $512*16*4KB=32MB$ in the 16-core BF3 SmartNIC. Considering that the 16 MB cache of BF3 is also shared with the TX path and other SmartNIC computations, the RX working set size would far exceed the available cache size.  
Figure~\ref{fig_cache_invalid}b demonstrates how a network stack receives packets when the working set exceeds the cache size. We observe that an incoming packet would always incur a cache eviction, thus, the system performance would be bottlenecked by the Arm memory bandwidth. 

\noindent\textbf{Unlimited-working-set in-cache processing RX. }
To this end, we propose unlimited-working-set in-cache processing RX, which avoids cache eviction even if the RX working set size far exceeds the cache size. Our key idea is based on the observation that the packet content is no longer needed by Arm after the network stack RX processes the packet header and forwards the packet to the host. As such, there is no need to actually evict the processed packets from the cache to the Arm memory. After the Arm core processes a packet and forwards the packet to the host, the Arm core would explicitly invalidate the cachelines of the packet buffer by invoking the BF3-provided API~\cite{cache-invalidate-operation}, which would set the related cachelines ``invalidate'' flag\footnote{We align each packet buffer to the cacheline granularity~(64B) to ensure that the invalidation would not affect other packets.}. 

Figure~\ref{fig_cache_invalid}c demonstrates how \sysname{} RX path receives packets when the working set exceeds the cache size. As the processed packet is explicitly invalidated, there are always available cachelines for incoming packets, and cacheline stale contents are discarded rather than written back, which allows the incoming packet to overwrite the cacheline directly and avoid unnecessary write back. In the rare case when a packet arrival burst occurs or the processing rate slows down, the incoming packets may find no invalidated cacheline to fill in and incur an eviction. However, the packet would not be dropped, and the eviction would disappear after the burst.

With unlimited-working-set in-cache processing RX, the required cache size only needs to be larger than the product of the network bandwidth and the average packet processing time. For a 400 Gbps network and a 10\us average network stack processing time, the required cache size is only 500 KB, which is far smaller than the BF3 Arm LLC size~(32$\times$ smaller). Even in the near future, when the network bandwidth scales to 800 Gbps and the processing latency is doubled, this mechanism still only needs 2MB LLC, comfortably within typical LLC capacities.

\begin{figure}
	\centering
	{\includegraphics[keepaspectratio=true, width=0.98\linewidth]{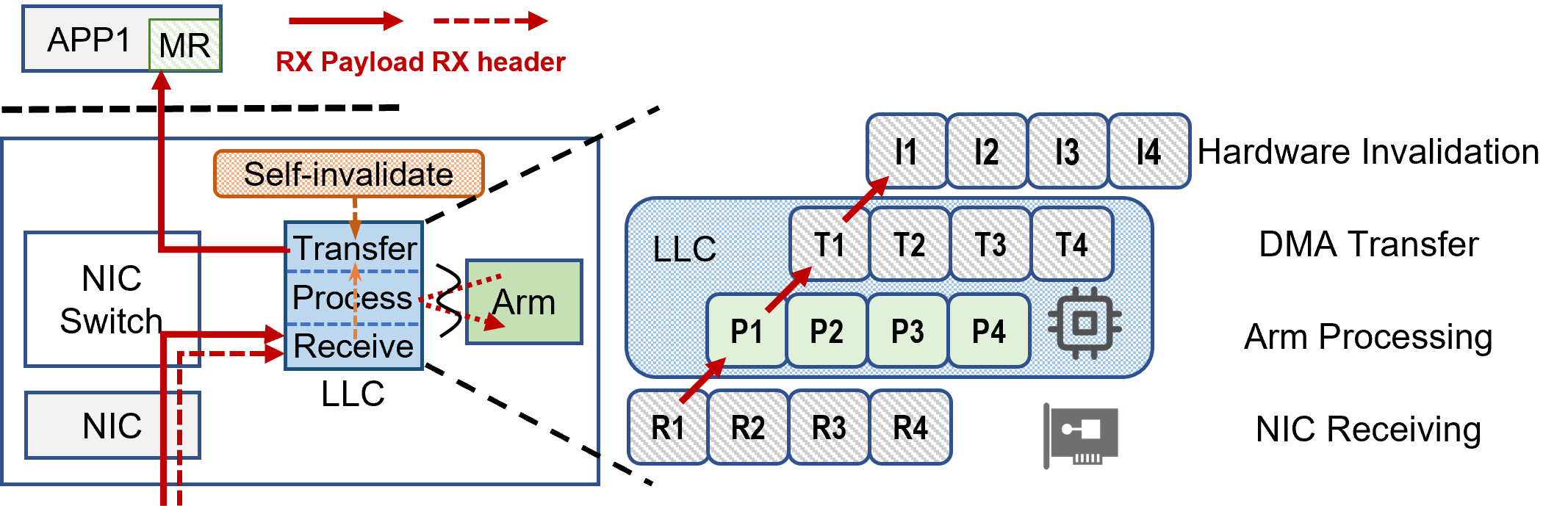}
        \vspace{-2ex}
	\caption{Architecture of unlimited-working-set in-cache processing execution pipe.}
	\label{fig_rx_pipe}}
        \vspace{-3ex}
\end{figure}

To further improve process performance, we decompose the RX path into four streaming stages that execute in a pipelined fashion as shown in Figure~\ref{fig_rx_pipe}: 1) The NIC receives an incoming packet and performs DMA to handoff it into the Arm LLC; 2) The corresponding Arm core processes the packet and sends an ACK to the sender; 3) The Arm core initiates the DMA engine to transfer the payload to the designated host buffer; 4) The Arm core explicitly self-invalidates the received packet buffer. Stage 1 is handled by NIC hardware, while stages 3 and 4 are executed asynchronously, consuming only a few Arm core cycles to launch. This pipeline is executed under a busy loop run-to-complete model, with each core dedicated to a specific RX queue to minimize inter-process communication and maximize cache locality. As a result, the Arm core is tasked solely with lightweight in-cache transport logic processing, while all heavyweight operations are delegated to the hardware.

\vspace{-2ex}
\subsection{DMA-only Notification Pipe}
\label{sec::dma_pipe}
\vspace{-1ex}

 Common RNICs utilize \textit{WQE-by-MMIO} and \textit{Doorbell} mechanisms to update SQ/RQ/CQ pointers with the RNIC~\cite{smart-dm, rdma-design-guide}. In the first case, the WQEs are transferred via 64B write-combined MMIOs; in the second case, the host updates the NIC Doorbell, followed by the NIC fetches WQEs using one or more DMAs. Although these mechanisms perform well for common RNIC, however, most off-path SmartNICs suffer limited capabilities emulated MMIO interface between the Arm and the host CPU~\cite{bf2,bf3}. Our observations reveal that BF3 achieves fewer than 1K/s MMIO write rate, which is markedly insufficient for the demands of 400Gbps networks. Moreover, exclusively using the Doorbell mechanism incurs additional latency due to extra PCIe round trips.

 
 Therefore, we propose a DMA-only notification pipe that relies solely on a high-performance DMA engine for communication and coordination. To maximize efficiency, the design enforces a single producer and a single consumer, ensuring lockless access and each element aligned to the cache line size. Each element contains a dedicated 1-bit flag signal to indicate its validity. The producer sets this flag in the producer buffer and initiates the DMA engine, while the consumer continuously polls the next element in the consumer buffer until its flag becomes valid. Upon reaching the buffer’s end, the flag toggles to indicate wrap‑around. Since the producer typically batches multiple elements per DMA transfer, cache contention is avoided~\cite{enso}. For SQ/RQ, the user application is the producer and the Arm is the consumer. To further realize memory alignment and keep cache locality, SQ elements (64B) are transferred immediately, whereas RQ entries are grouped in batches of four (4 $\times$ 16B). For the CQ element (64B), roles are reversed: the Arm produces entries, and the user application consumes them. But it involves a question about how the Arm tracks the user application’s progress in handling CQE. To address this, we implement a consumer counter located in the user library: each time the consumer processes an element, it automatically increases the counter, and the producer periodically reads it via one DMA read after every $n$ elements (customized by developers).

\noindent\textbf{Latency-sensitive flow optimization.} 
We observe that latency-sensitive applications typically have payload sizes limited to a single MTU and employ two-sided operations. To address challenge \#3 (\textbf{C3}), we introduce a low-latency QP optimized for MTU‑sized, two‑sided operations, comprising two complementary mechanisms: In the TX path, we extend the SQE to carry inline payloads, thereby eliminating one PCIe round‑trip. Payloads are delivered to the Arm via our high‑performance notification pipe followed by inline send~\cite{rdma-design-guide}. In the RX path, we support the NIC directly placing the payload section to the application buffer while routing the header section to the Arm. The application polls the first byte of the receive buffer to detect incoming packets. In this case, the transport protocol is still processed on the SmartNIC, while the application should detect and handle out‑of‑order and duplicate packets at the user level.

\vspace{-2ex}
\subsection{Programmable Offloading Engine}
\label{sec:one_side}
\vspace{-1ex}

In addition to the network stack, exploiting the spare SmartNIC computing resources to mitigate the host CPU occupation is a trendy way to lower the data center tax~\cite{linefs_sosp21,ipipe,fuyao}. However, prior works~\cite{RPCAcc, cdpu,benchbf,smartds,strom} require ad-hoc design and deployment tailored to specific scenarios and applications. \sysname{} propose a programmable offloading engine, allowing easy offloading of various applications. The engine runs on top of the network stack transport and provides a rich set of programming APIs to allow cloud providers to offload the application logic easily. In short, programmable offloading engine allows cloud providers to register an unused transport opcode and hook up with a customized function that runs on the spare SmartNIC Arm cores.



This offloading engine executes application-layer functions on dedicated Arm cores isolated from the network stack and can invoke lookaside accelerations on the SmartNIC to provide computing intensity. Cloud providers configure the number of Arm cores allocated to this engine at \sysname{} startup. When the network stack receives a packet bearing a registered opcode, it processes this packet as a normal ``SEND'' packet, transfers the payload to the pre-posted receive buffer, and performs cache invalidation. After receiving, the network stack forwards it to the offloading engine via the atomic queue~(akin to the DMA-only notification pipe but implemented using atomic load/store instructions), whereupon the engine invokes the registered handler for further processing.


\vspace{-1ex}
\begin{table} [htbp]
	\centering
 	\caption{Programmable offloading engines APIs.}
	\label{tab_api}	
        \vspace{-2ex}

	\begin{tabular}{|l|}
		\hline
            \makecell[l]{\textbf{register\_opcode(opcode, qp, func)}}\\
            \hline
            \makecell[l]{
                Registering the target opcode to the network stack, this \\ function will be invoked when a target packet is received.
            }\\
            \hline
            \hline

		\hline
            \makecell[l]{\textbf{register\_dma\_region(host\_addr, size)}}\\
            \hline
            \makecell[l]{
                Registering the host memory on the Arm for \\ following DMA operations.
            }\\
            \hline
            \hline
    
		\hline
            \makecell[l]{\textbf{alloc\_resp(context, size)}}\\
            \hline
            \makecell[l]{
                Allocating $size$ bytes on pinned Arm memory for \\ response packet, and return the address point.
            }\\
            \hline
            \hline

            \makecell[l]{\textbf{submit\_dma(context, op, host\_addr, arm\_addr, size)}}\\
            \hline
            \makecell[l]{
                Submit a DMA operation to the associated address, op \\ can be READ or WRITE, return an ID for trace.
            }\\
            \hline
            \hline

            \makecell[l]{\textbf{wait\_dma\_finish(context, dma\_id)}}\\
            \hline
            \makecell[l]{
                Wait the dma\_id corresponding dma operation to finish.
            }\\
            \hline
            \hline

            \makecell[l]{\textbf{submit\_resp(context, addr, size)}}\\
            \hline
            \makecell[l]{
                Submit the response packet with the specific address.
            }\\
            \hline
	\end{tabular}
        \vspace{-2ex}
\end{table}

This offloading engine provides a rich set of programming API~(Table~\ref{tab_api}) to help cloud providers deploy their offloading tasks. As such, cloud providers can focus on task logic instead of dealing with low-level network stack details. We take a simple batched RDMA READ as an example and show how to use these APIs to implement it; the complete source code is provided in Appendix $\S$\ref{codeing_example}. The programmer first chooses an unused opcode and uses \texttt{register\_opcode} to register a batched RDMA READ handler function. Before establishing the QP connection, the programmer invokes \texttt{register\_dma\_region} to register a host memory region for subsequent DMA accesses. Upon receiving a corresponding request with the batched RDMA READ opcode, that registered handler function is invoked and executed as a user-space coroutine. In this coroutine, \texttt{alloc\_resp} is called to allocate a pinned Arm memory for the response packet. The programmer then invokes \texttt{submit\_dma} to execute the DMA operation, and all DMA‑related tasks are enqueued into a task pool and executed asynchronously via coroutines; once the DMA transfer completes, \texttt{wait\_dma\_finish} returns control and execution resumes. Finally, the programmer uses \texttt{submit\_resp} to deliver the response packet back to the client.

\vspace{-2ex}
\section{Implementation}
\vspace{-2ex}

We build a fully functional prototype of \sysname{} using Nvidia BlueField-3 SmartNIC~\cite{bf3} with a PCIe5.0$\times$16 interface and 2$\times$200 Gbps Ethernet ports. \sysname’s implementation consists of the core network stack running as a separate user process on the Arm processor, the kernel modules running in the host kernel, and the user-space libraries linking with user applications. The core network stack is implemented in 6,890 lines of C++20 code, the kernel module in 1,350 lines of C98 code, and the user-space libraries in 3,012 lines of C++20 code. Moreover, we introduced several modifications (500 lines of code) to the mlx5 driver on the Arm to enable more efficient DMA and cache operation interfaces, while \textbf{keeping the host mlx5 driver unchanged}. Our entire system runs in an unmodified Linux environment. We left more implementation details in Appendix $\S$\ref{implement_details}.

\vspace{-2ex}
\section{Evalution}
\vspace{-2ex}

Our evaluations aim to answer the following questions:
\squishlist
    \item How does the performance of \sysname{} compare to other network stacks ($\S$\ref{evaluation_overview})?
    \item How effective is the header-only offloading TX path ($\S$\ref{evaluation_tx})?
    \item How effective is the unlimited-working-set in-cache processing RX path ($\S$\ref{evaluation_rx})?
    \item How effective is the DMA-only notification pipe ($\S$\ref{evaluation_notification_pipe})? 
    \item How effective is the programmable offloading engine ($\S$\ref{evaluation_smart_one_side})?
    \item How much performance acceleration can \sysname{} achieve for block storage and KVCache transfer workloads ($\S$\ref{evaluation_e2e})?
\squishend

\vspace{-2ex}
\subsection{Experimental Setup}
\vspace{-1ex}

\noindent{\bf Hardware Testbed.} 
Our hardware testbed consists of two servers, each having two 16-core Intel Xeon Gold 6426Y running at 2.5GHz, 512 GiB (16x32 GiB) 4800 MHz DDR5 memory, and a 37.5 MiB LLC. Each server is equipped with an Nvidia BlueField-3 B3220 400GbE NIC and connected back-to-back using two 200GbE QSFP56 cables.

\noindent{\bf Snap Baseline.} 
We implement the baseline "Snap" that runs the network stack as a separate user program and dedicated CPU cores. As a representative of microkernel-based network stacks, Snap leverages DPDK~\cite{DPDK} to achieve high throughput and adopts a simple go-back-N mechanism for packet loss recovery. Since Snap~\cite{Snap} is not open-sourced, we construct our network stack based on the descriptions provided in the paper while minimizing any unnecessary CPU overhead.

\noindent{\bf RDMA NIC Baseline.}
We implement the baseline "RNIC" that directly uses the hardware-offloaded RDMA stack provided by the state-of-the-art RNIC Nvidia ConnectX-7~\cite{cx7}, leveraging RC mode and RoCEv2 protocol to achieve the highest performance. We use dedicated busy-looping CPU cores to execute IBV verbs like \texttt{post\_send} and \texttt{poll\_cq}. 

\noindent{\bf Solar-CPU Baseline.} 
We implement the Solar~\cite{solar} transport protocol on dedicated host CPU cores, strictly following the specifications outlined in the paper. Solar is the storage network stack for Alibaba Cloud's EBS service and has been deployed on a large scale. Due to the current lack of commodity RNIC support for the Solar protocol, we deploy it on the host CPU and leverage CRC hardware offload along with DSA engines to achieve optimal performance.

\vspace{-2ex}
\subsection{Comparison with other Network Stacks}
\label{evaluation_overview}
\vspace{-1ex}


\begin{figure}[]
    \subfloat[SEND throughput, TX depth=64]{
        \label{fig_e_comp_send_throughput}
        \includegraphics[keepaspectratio=true, width=0.50\linewidth]{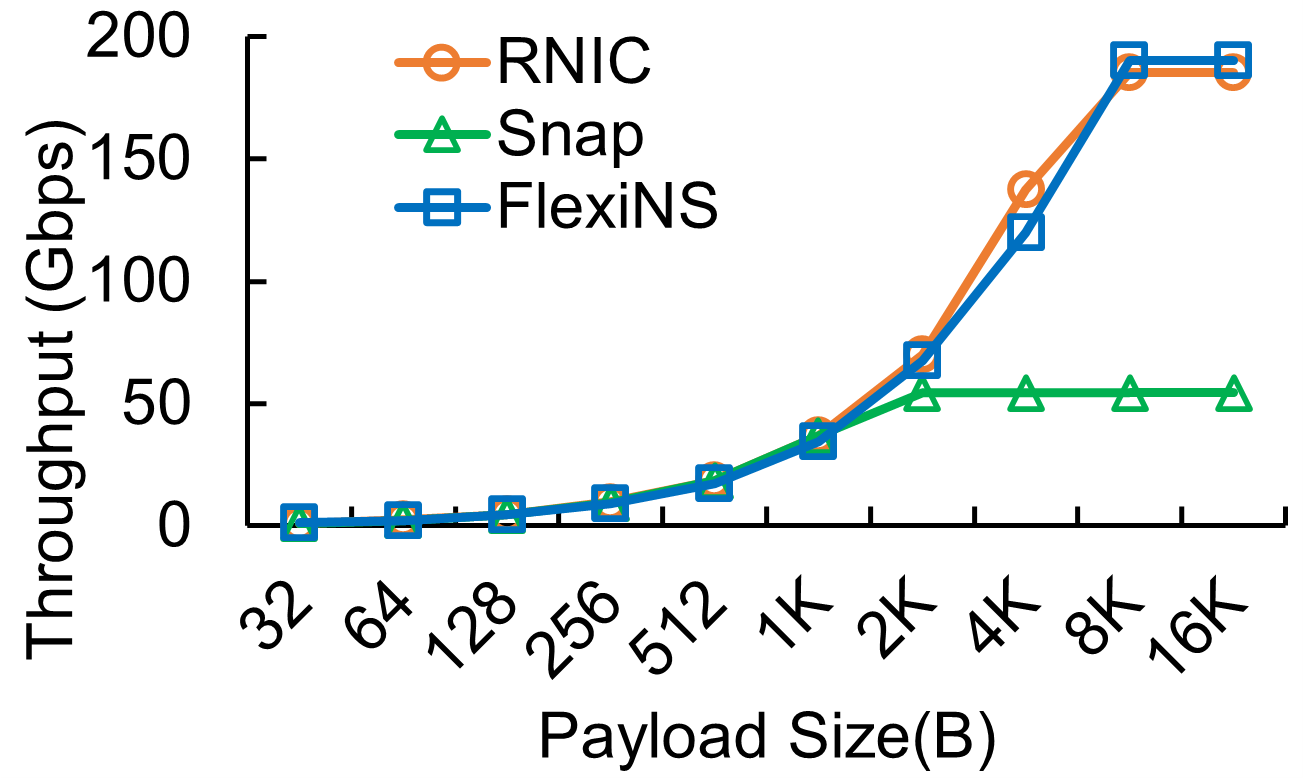}
    }
    \subfloat[WRITE throughput, TX depth=64]{
        \label{fig_e_comp_write_throughput}
        \includegraphics[keepaspectratio=true, width=0.50\linewidth]{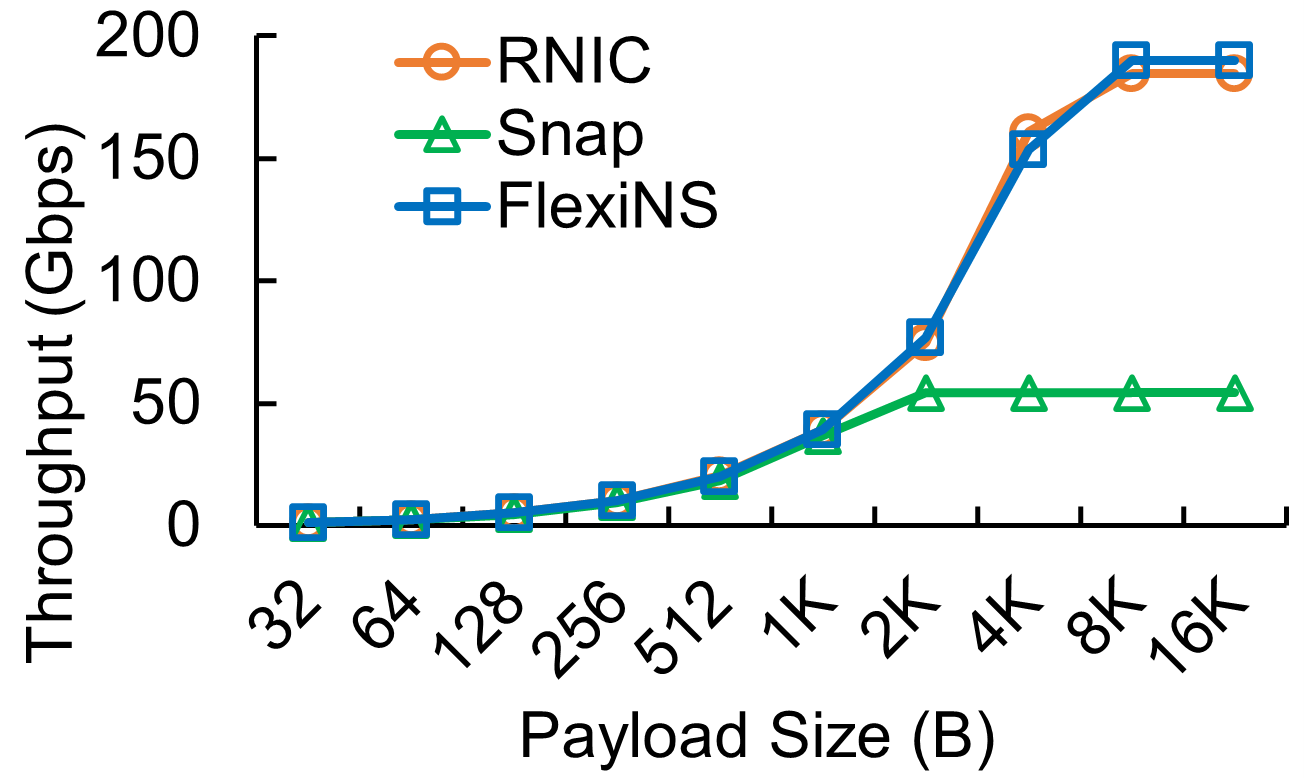}
    }
    \vspace{-2ex}
    \newline
    \subfloat[SEND latency, TX depth=1]{
        \label{fig_e_comp_send_latency}
        \includegraphics[keepaspectratio=true, width=0.50\linewidth]{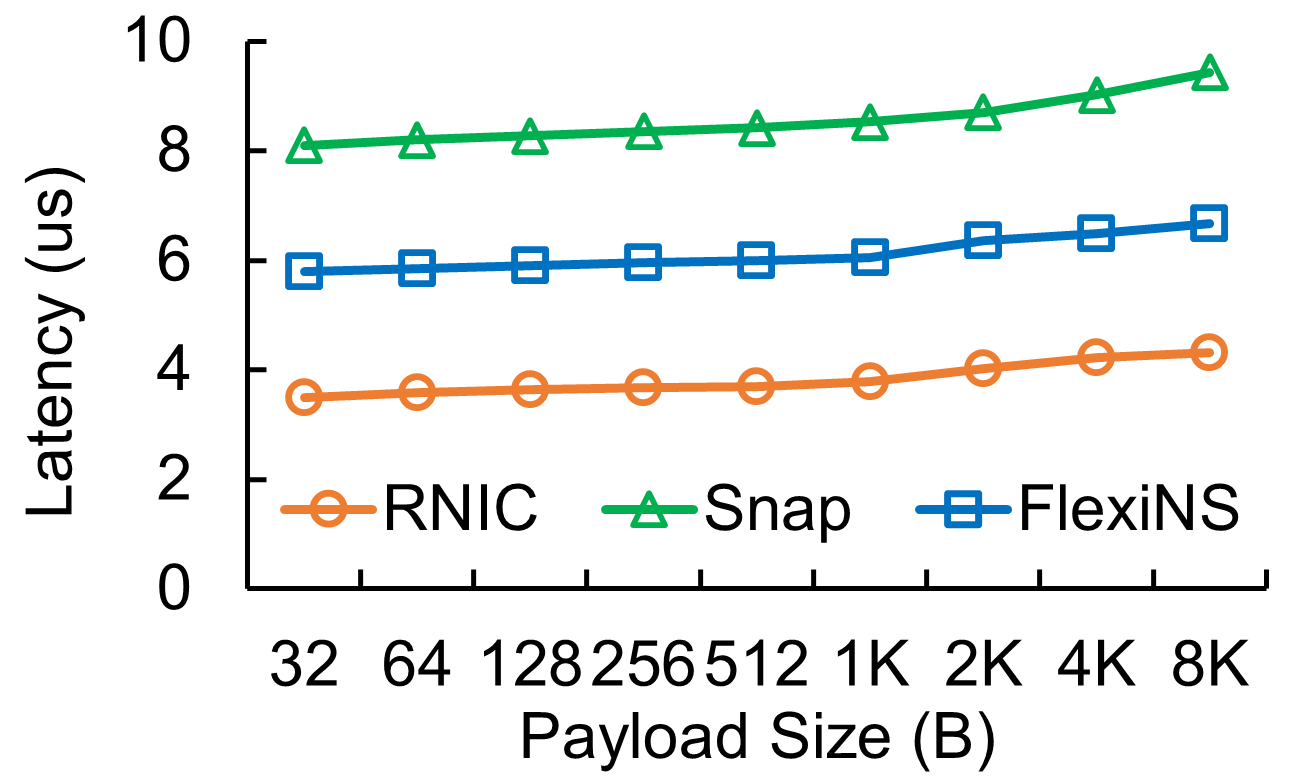}
    }
    \subfloat[WRITE latency, TX depth=1]{
        \label{fig_e_comp_write_latency}
        \includegraphics[keepaspectratio=true, width=0.50\linewidth]{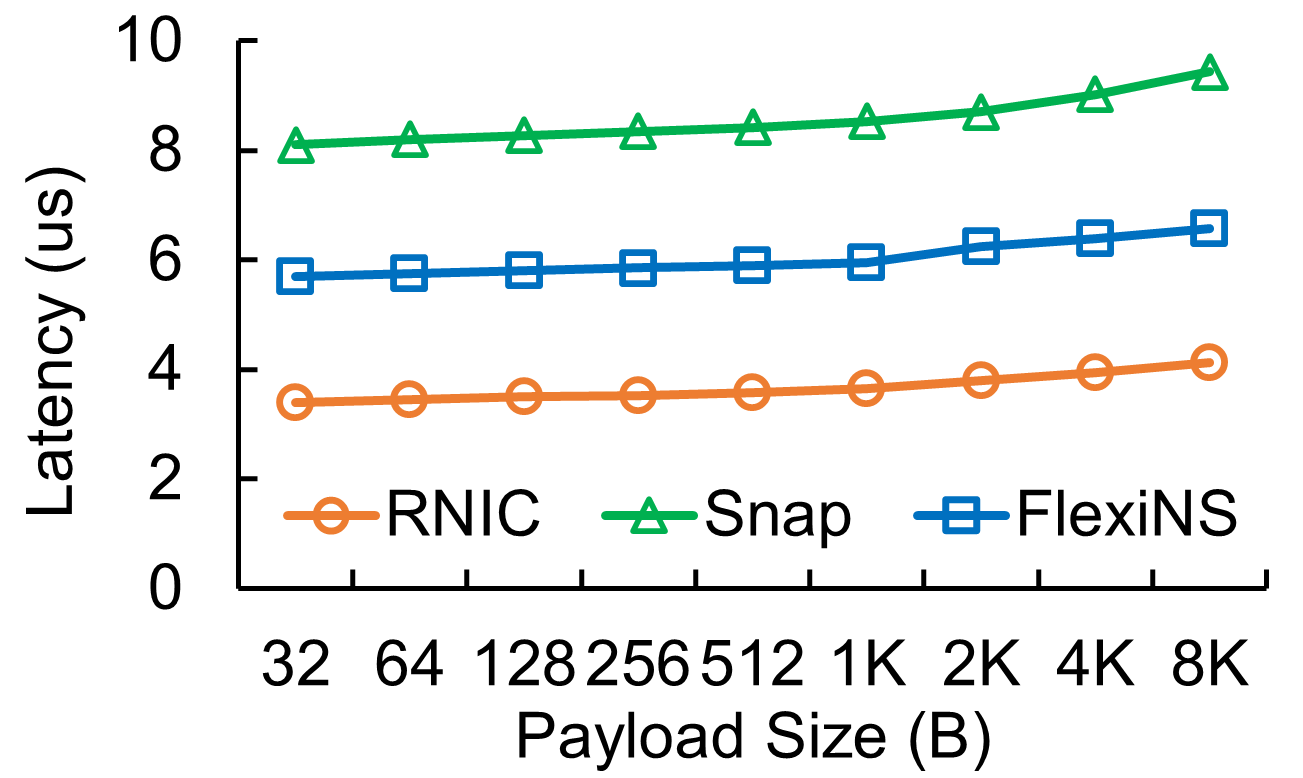}
    }
    \vspace{-1ex}
    \caption{RDMA SEND/WRITE throughput and latency between a pair of connections on different hosts.}
    \label{fig_e_comp_send_write}
    \vspace{-4ex}
\end{figure}

In this section, we focus on two basic performance metrics: throughput and latency. We use the LibR benchmark tool~\cite{libr2025}, which is similar to perftest~\cite{perftest2025} but has finer-grained control.
These tools can run on \sysname{} without any modification, and the dedicated CPU cores assigned to the network stack remain in a spin state to maximize performance.

Figure~\ref{fig_e_comp_send_write} illustrates the single connection throughput and latency of the SEND/WRITE operations across various network stacks. We set TX depth to 64 for the throughput test and 1 for the latency test. We have two observations. First, \sysname{} achieves comparable throughput to RNIC and up to 3.5$\times$ higher throughput than Snap in both SEND and WRITE tests. This is because \sysname{} offloads massive stack workloads to NIC hardware similar to RNIC, but Snap relies on the host CPU to execute, constraining single-connection throughput. Second, \sysname{} exhibits 1.5× higher latency than RNIC but still 1.4$\times$ lower than Snap, because WQE and CQE in \sysname{} must cross the Arm-NIC switch link and suffer PCIe interconnect latency relative to RNIC. 
However, the proximity of the Arm to the NIC enables \sysname{} to reduce network latency compared to Snap. 

\begin{figure}[]
    \subfloat[Aggreagate WRITE throughput]{
        \label{fig_e_comp_multi_qp_throughput}
        \includegraphics[keepaspectratio=true, width=0.50\linewidth]{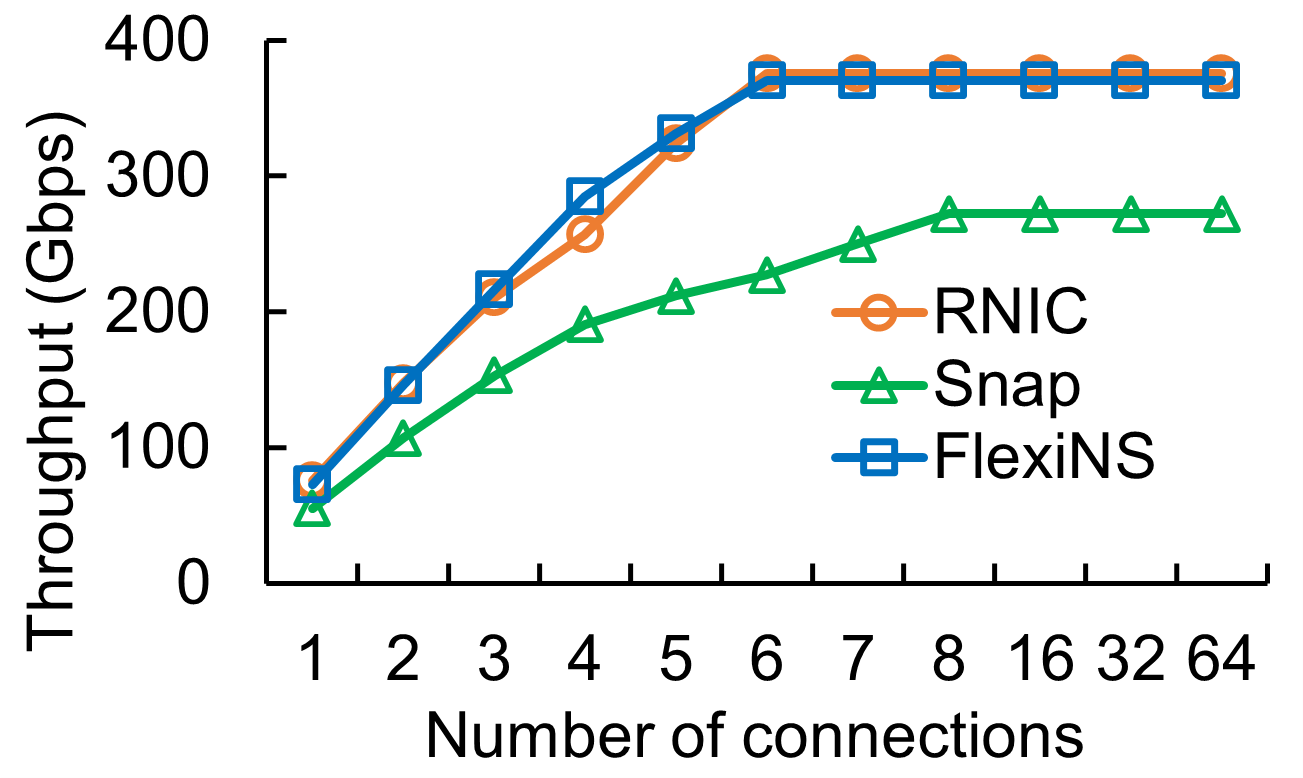}
    }
    \subfloat[Host memory bandwidth occupation]{
        \label{fig_e_comp_multi_qp_memory_bw}
        \includegraphics[keepaspectratio=true, width=0.50\linewidth]{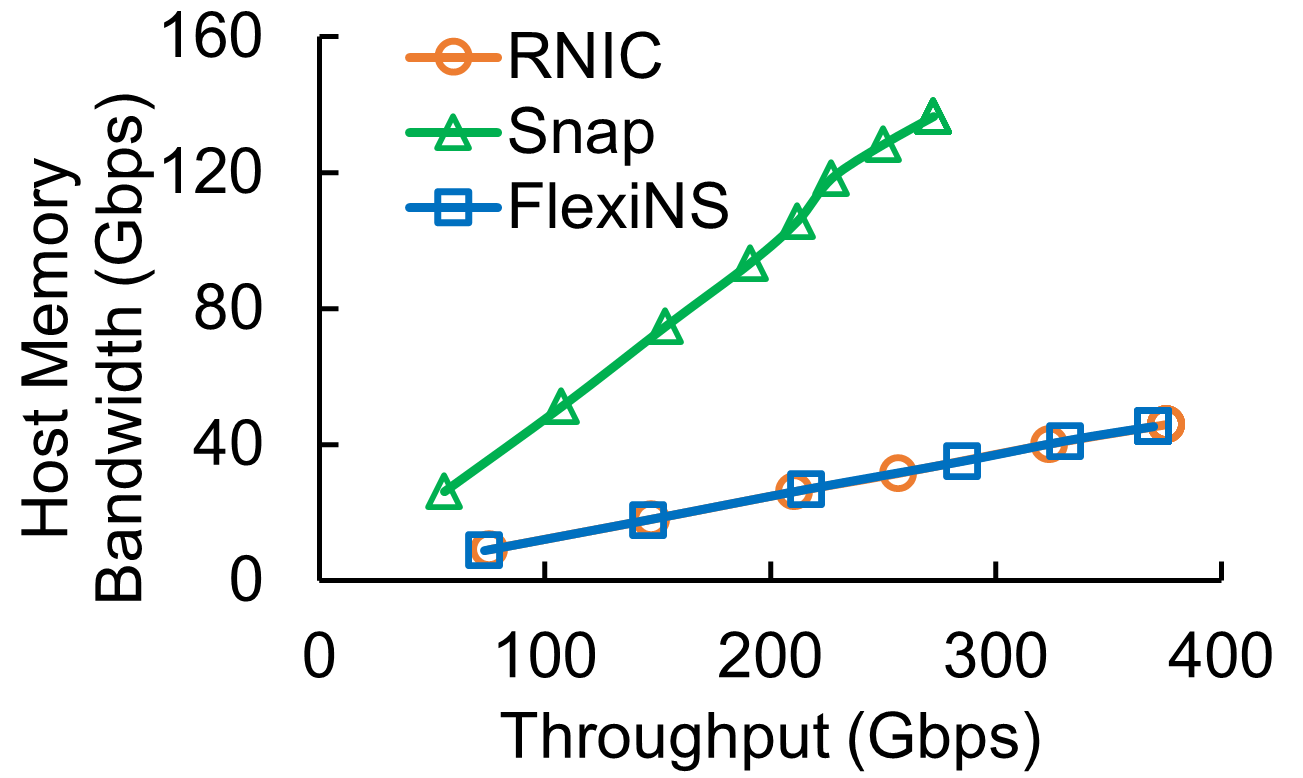}
    }
    \vspace{-1ex}
    \caption{Aggreagate RDMA WRITE throughput of multiple connections and corresponding host memory bandwidth.}
    \label{fig_e_comp_multi_qp}
    \vspace{-4ex}
\end{figure}

Figure~\ref{fig_e_comp_multi_qp} shows the throughput of WRITE operations with multiple connections, and the corresponding host memory bandwidth across various network stacks, where the TX depth of each connection is set to 64 and payload size is set to 2KB. We observe that \sysname{} demonstrates nearly linear scaling with the increase in connections, reaching line rate with more than 5 connections, comparable to RNIC and 1.4$\times$ higher than Snap. \sysname{} also keeps minimal host memory bandwidth occupation and reduces the 2.7$\times$ bandwidth compared with Snap. This efficiency is attributed to its SmartNIC-based packet handling and zero-overhead on the host CPU.

\vspace{-2ex}
\subsection{Header-only Offloading TX Path}
\label{evaluation_tx}
\vspace{-1ex}

In this section, we examine the performance of our header-only offloading TX path. As the baseline, we implement the na\"ive RDMA-assisted entirely offloading TX and DMA-assisted entirely offloading TX approaches discussed in ($\S$\ref{sec:tx_path}). Moreover, since BF3 has two ports and each connection only uses one port, we use two connections and set TX depth to 64, finally we collect and count the aggregated bandwidth. 

\begin{figure}[]
    \subfloat[WRITE throughput, TX depth=64]{
        \label{fig_e_tx_throughput}
        \includegraphics[keepaspectratio=true, width=0.50\linewidth]{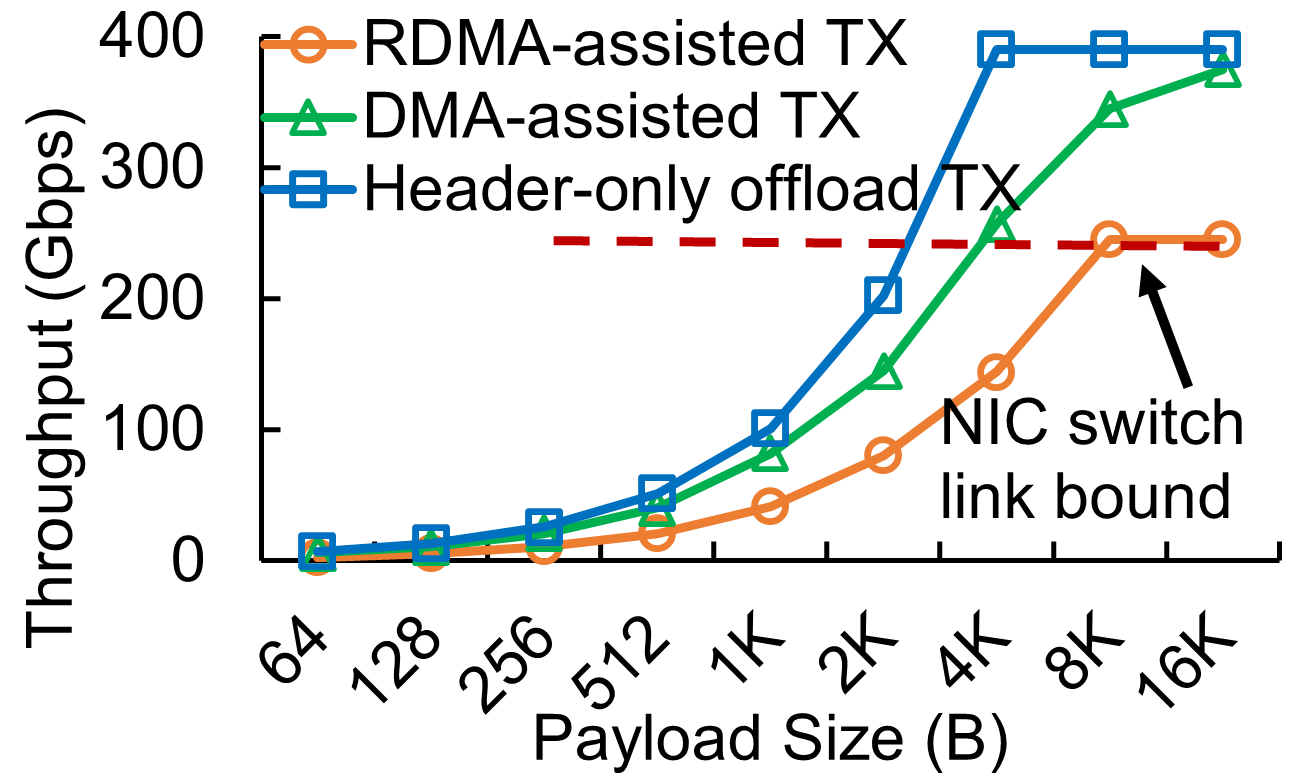}
    }
    \subfloat[Arm memory bandwidth occupation]{
        \label{fig_e_tx_arm_memory}
        \includegraphics[keepaspectratio=true, width=0.50\linewidth]{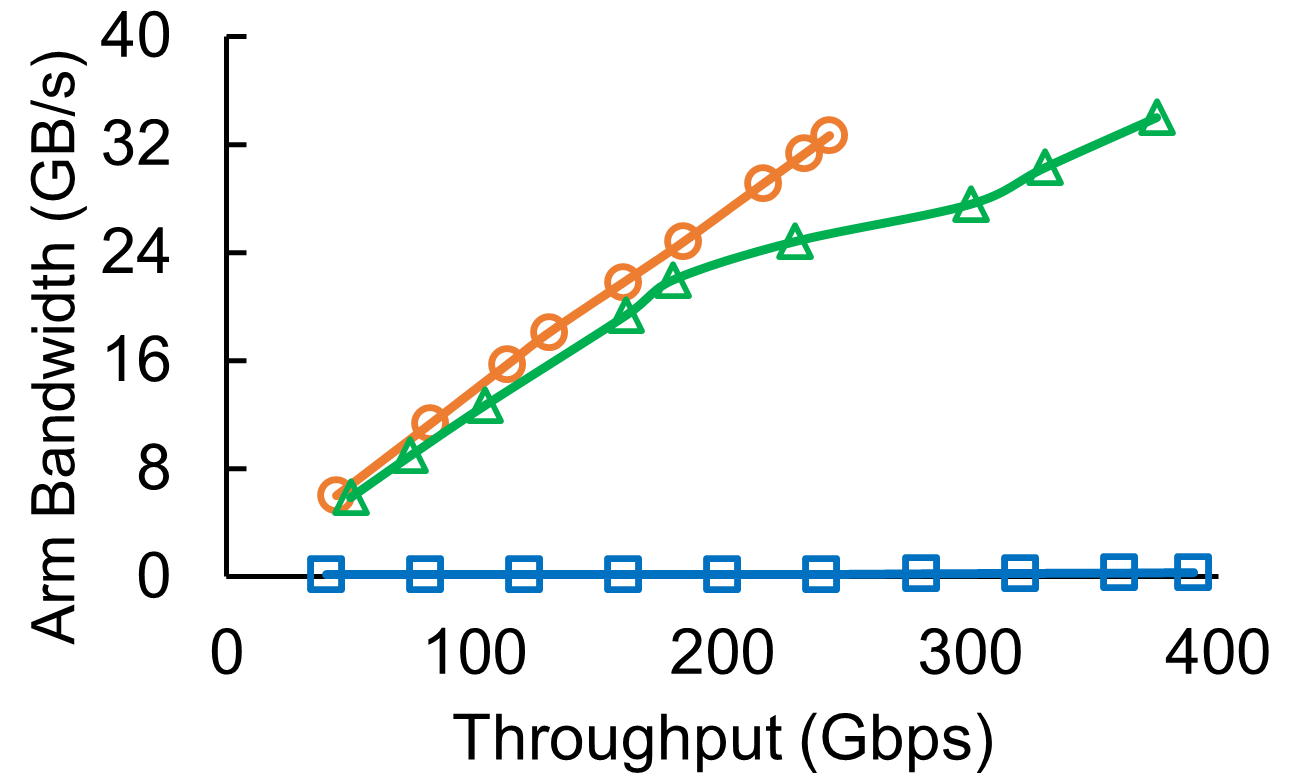}
    }
    \vspace{-1ex}
    \caption{Comparison of RDMA WRITE throughput under 2 connections across three TX Path choices and corresponding Arm memory bandwidth.} 
    \label{fig_e_tx}
    \vspace{-3ex}
\end{figure}

Figure~\ref{fig_e_tx_throughput} shows the different TX approaches achieved throughput, and Figure~\ref{fig_e_tx_arm_memory} illustrates the corresponding Arm memory bandwidth occupation. We have three observations. First, when the payload size is below 8KB, header-only offloading TX achieves 2.7$\times$ and 1.5$\times$ higher throughput, because this mechanism only needs to construct the packet header and achieve zero copy, and thus significantly reduces the Arm core and memory subsystem pressure. Second, header-only offloading TX and DMA-assisted entirely offloading TX reach line rate, whereas RDMA-assisted entirely offloading TX is constrained by Arm-NIC switch link utilization, as discussed in Section ($\S$\ref{sec:tx_path}). Third, header-only offloading TX maintains Arm memory usage below 0.5GB/s, independent of network throughput and 70$\times$ lower than DMA-assisted TX. This efficiency is achieved by the header-only offloading TX bypassing the packet payload storage. 

\begin{figure}
	\centering
	{\includegraphics[keepaspectratio=true, width=0.9\linewidth]{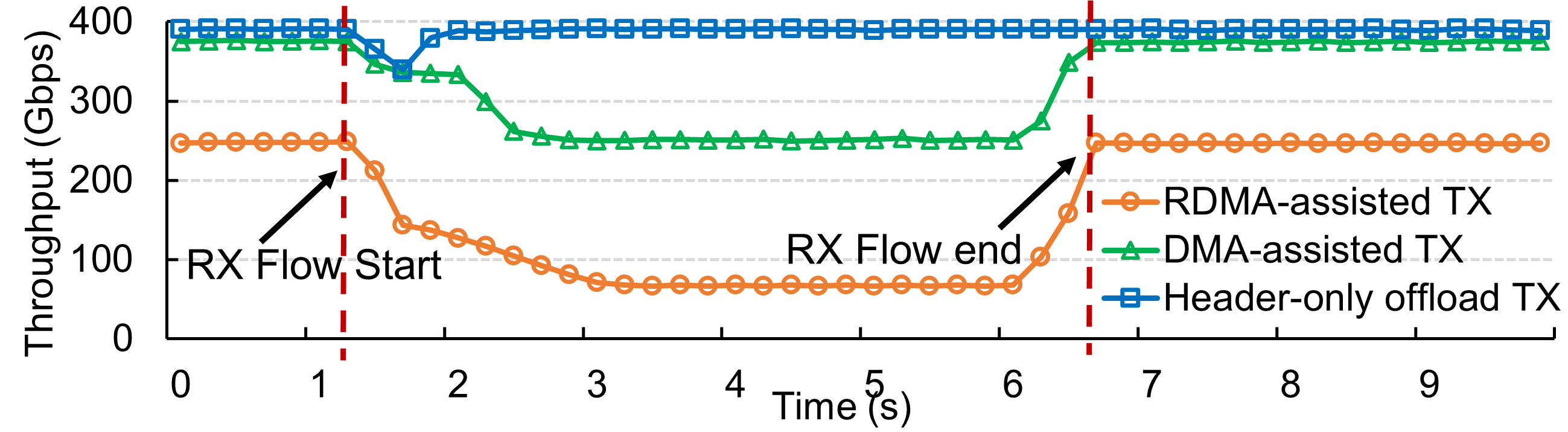}
        \vspace{-1ex}
	\caption{Using 8 connections and 2KB RDMA WRITE payload to achieve each TX design upper limit, and insert a 400Gbps RX flow at the 1 second that lasts 5 seconds.}
	\label{fig_e_tx_reversed_flow}}
    \vspace{-4ex}
\end{figure}

To further illustrate the benefits, we use 8 connections and 2KB payload to repeat the above experiment. Once throughput stabilized at its maximum, we insert a 400 Gbps RX flow on different cores from the server Arm core to the client Arm core, running for 5 seconds. Figure~\ref{fig_e_tx_reversed_flow} shows the aggregate TX throughput achieved by the various approaches. Notably, our approach maintains line rate despite the RX flow, whereas the other two experience 72\% and 35\% throughput reduction. This is because our mechanism avoids contention on the Arm-NIC switch link with the RX flow.

\vspace{-2ex}
\subsection{Unlimited-working-set In-Cache \\ Processing RX Path}
\label{evaluation_rx}
\vspace{-1ex}

In this section, we evaluate the effectiveness of the unlimited-working-set in-cache processing RX path. We use 12 Arm cores on both client and server, each core handles a single connection. We equip each RX queue with 8KB queue element to receive jumbo packets. On the server, we vary the number of RX queue elements to emulate different working set sizes. The client sends 8KB RDMA WRITE requests, which are processed by the server's Arm cores, and then the payload is transferred to the host buffer by inter-node RDMA/DMA. Similar to the TX path, we compare three strategies: (1) using na\"ive intra-node RDMA for transferring payload from Arm to Host, (2) using na\"ive DMA for transferring, and (3) using unlimited-working-set in-cache processing RX. 

\begin{figure}[]
    \subfloat[WRITE throughput, 8KB payload]{
        \label{fig_e_rx_throughput}
        \includegraphics[keepaspectratio=true, width=0.50\linewidth]{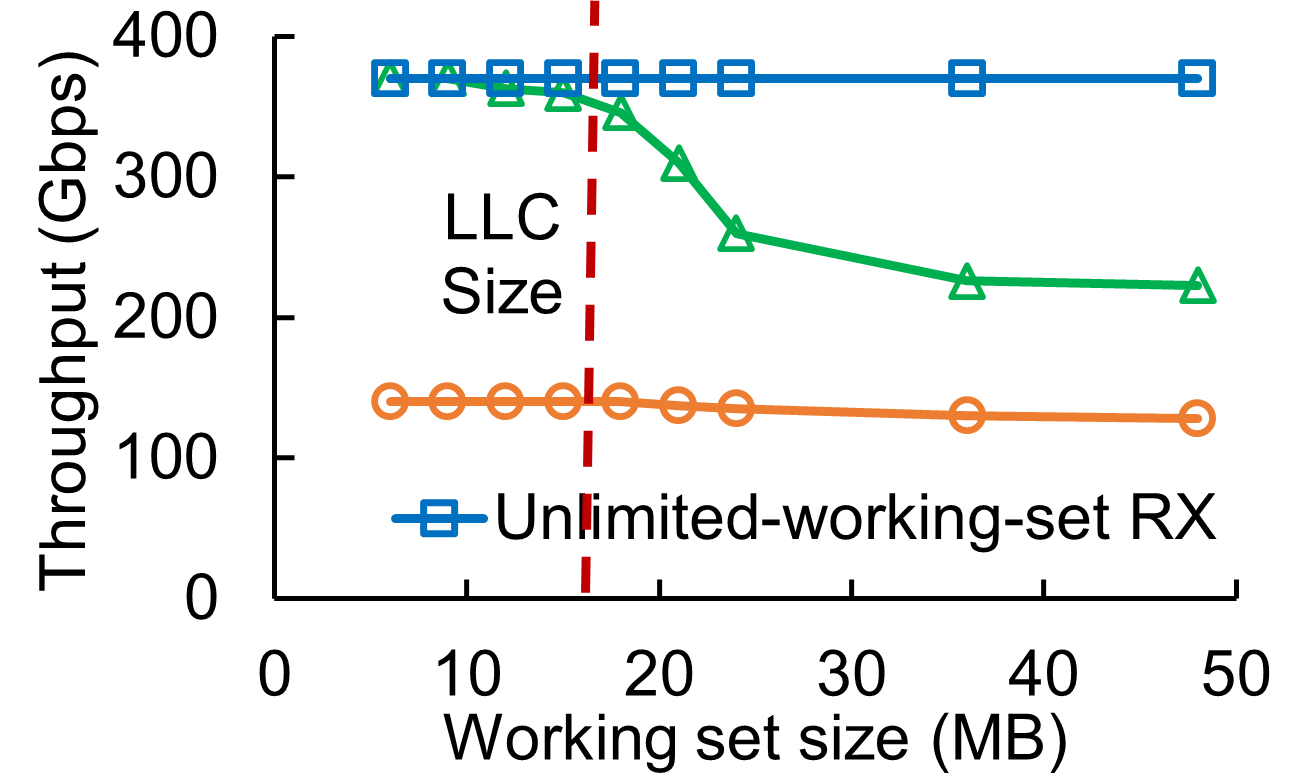}
    }
    \subfloat[Arm memory bandwidth occupation]{
        \label{fig_e_rx_arm_memory}
        \includegraphics[keepaspectratio=true, width=0.50\linewidth]{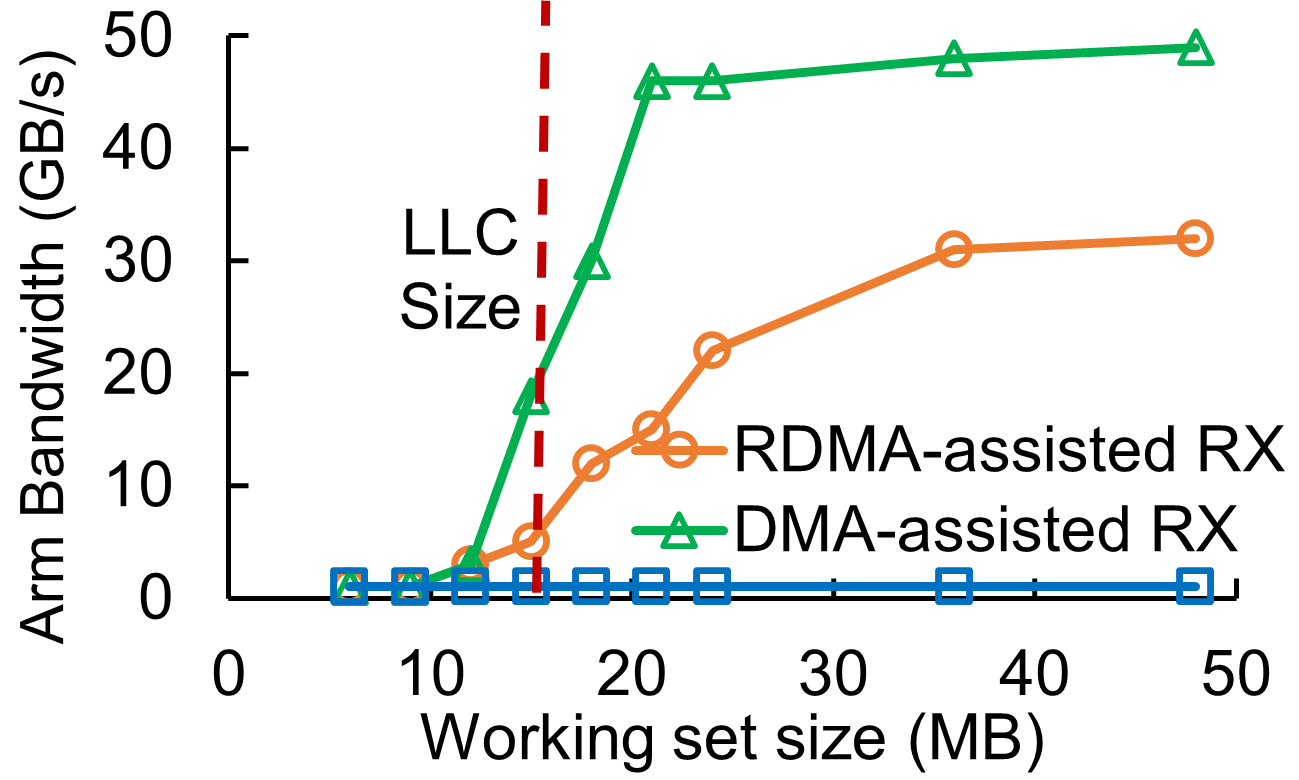}
    }
    \vspace{-1ex}
    \caption{Comparison of 8KB RDMA WRITE throughput under different working set sizes across three RX path choices and corresponding Arm memory bandwidth.}
    \label{fig_e_rx}
    \vspace{-4ex}
\end{figure}

Figure~\ref{fig_e_rx_throughput} shows the different RX working set sizes achieved throughput, while Figure~\ref{fig_e_rx_arm_memory} illustrates the corresponding Arm memory bandwidth. We have two observations. First, the unlimited‐working‐set mechanism maintains line‐rate throughput even as the working set far exceeds LLC size; specifically, \sysname{} achieves 1.6$\times$/2.9$\times$ higher throughput than na\"ive DMA/RDMA transfer when the working set size is above 48MB, primarily due to a significant reduction in pressure on the Arm memory subsystem. Notably, even under 48MB working set size, each RX queue only contains fewer than 512 elements, which is relatively small. Second, the unlimited-working-set mechanism in-cache processing RX introduces less than 0.8GB/s of memory bandwidth usage even when throughput reaches line rate, demonstrating that nearly all operations (receive, process, and transfer) are processed in cache, without incurring additional memory bandwidth overhead. Third, the na\"ive RDMA/DMA exhibits substantial Arm bandwidth utilization once the working set size exceeds the LLC, indicating a pronounced leaky DMA problem effect that limits achievable throughput.

\vspace{-2ex}
\subsection{DMA-only Notification Pipe}
\label{evaluation_notification_pipe}
\vspace{-1ex}

\begin{figure}[]
    \subfloat[Throughput and latency of \\ different notify mechanisms]{
        \label{fig_e_notifaction_pipe}
        \includegraphics[keepaspectratio=true, width=0.50\linewidth]{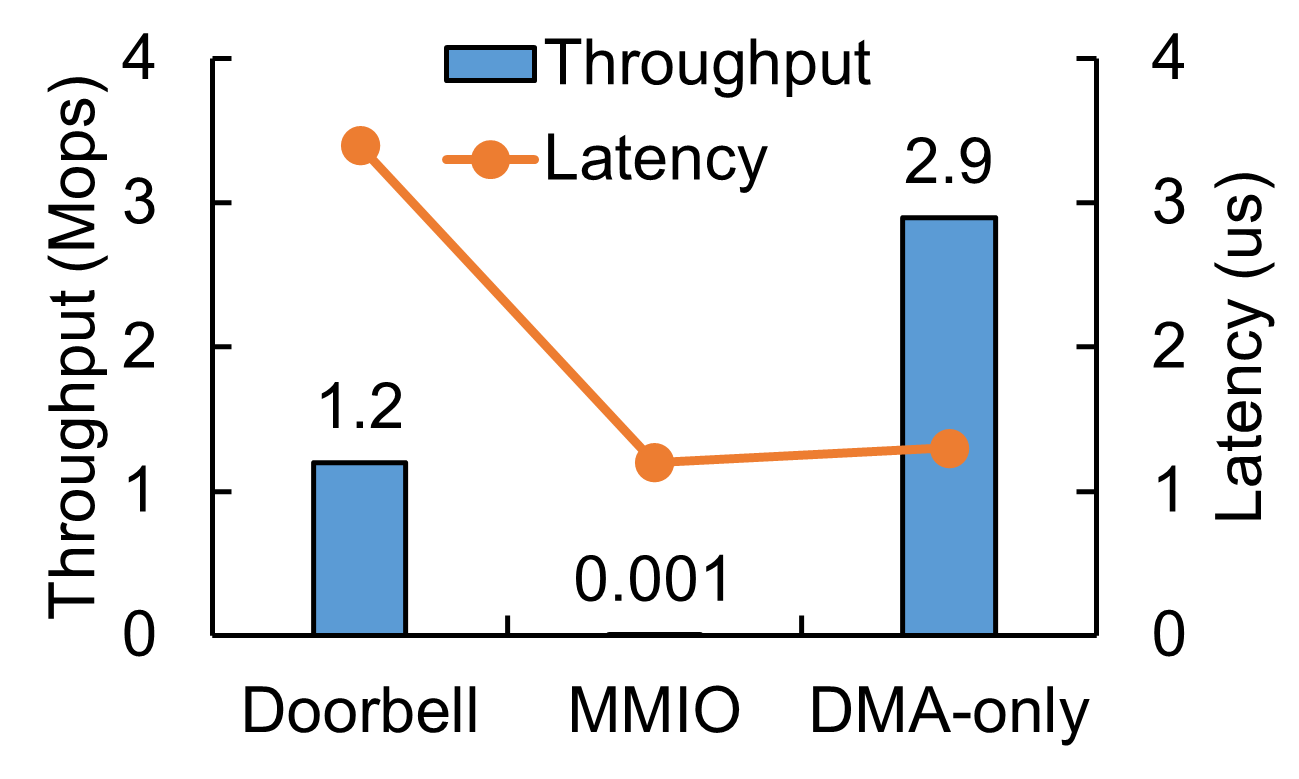}
    }
    \subfloat[L2 reflector latency for \\ different network stack]{
        \label{fig_e_low_latency_qp}
        \includegraphics[keepaspectratio=true, width=0.50\linewidth]{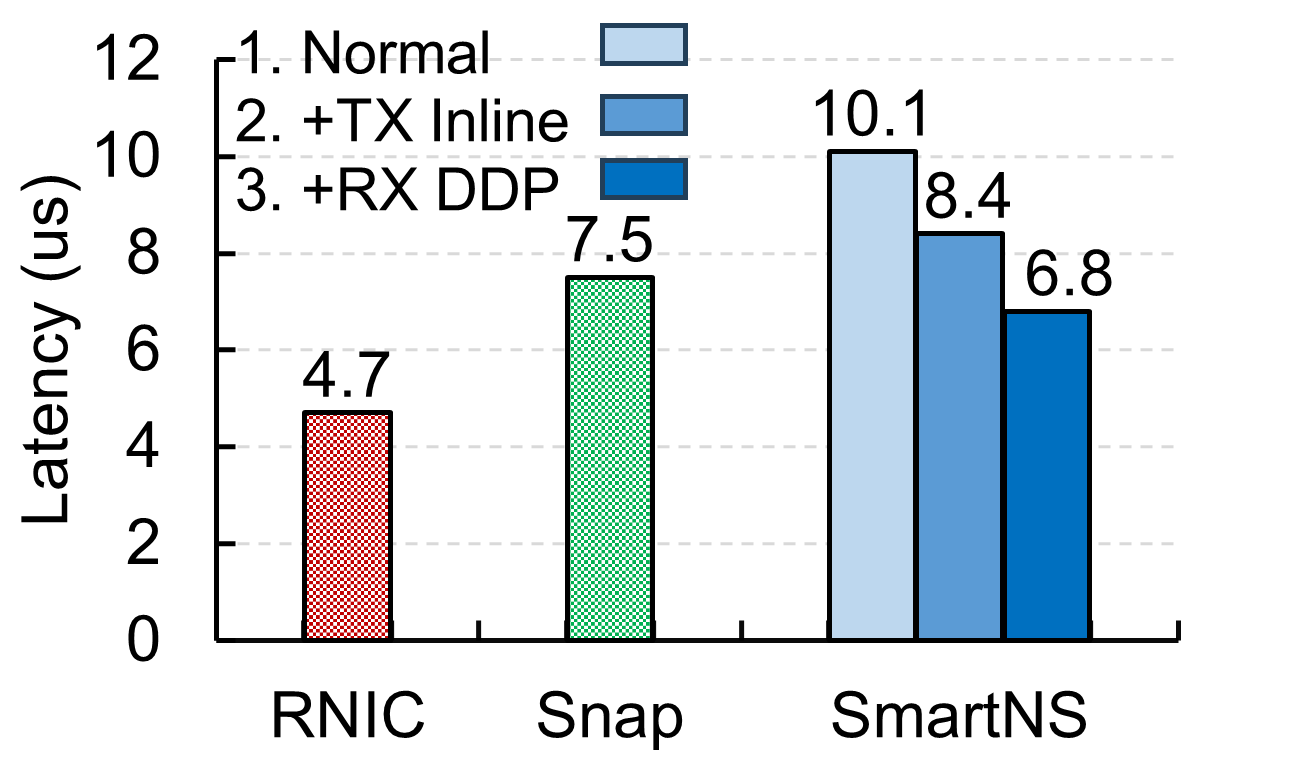}
    }
    \vspace{-1ex}
    \caption{Performance of DMA-only notification pipe and low-latency QP.}
    \label{fig_e_dma_only_low_latency}
    \vspace{-4ex}
\end{figure}

In this section, we compare \textit{Doorbell}, \textit{WQE-by-MMIO}, and DMA-only notification mechanisms for WQE submission latency and achieved throughput. Here, we define latency as the elapsed time between host submission of a 64‑byte WQE and its receipt by the Arm; both tests employ one host CPU and one Arm core. Figure~\ref{fig_e_notifaction_pipe} shows the throughput and latency of different notify mechanisms. We have two observations. First, DMA-only notification pipe achieves comparable latency with MMIO, and 2.6$\times$ lower than Doorbell, this is mainly because it omits the additional PCIe round-trip. Second, it delivers the highest throughput, which is 2.4$\times$ higher than the Doorbell mechanism; this is also caused by the extra PCIe round trip. Surprisingly, the MMIO mechanism exhibits such low throughput, less than 1K/s. This is because most off-path SmartNICs suffer limited capabilities emulated MMIO interface, the Arm must communicate with NIC firmware to get the content of incoming requests, which is very costly.

We also evaluate the latency of different network stacks via an L2-reflector application. The client sends a 64-byte packet to the server, and the server swaps the source and destination MAC addresses of each packet and returns it to the client. As shown in Figure~\ref{fig_e_low_latency_qp}, unoptimized \sysname{} suffers 2.2$\times$/1.4$\times$ higher latency than RNIC and Snap, respectively. However, with our TX inline and RX direct data placement optimization, \sysname{} can achieve 1.11$\times$ lower latency than Snap. Although it is still approximately 2\us slower than RNIC, this margin is modest. Consequently, developers can freely choose \sysname{} for its programmability or RNIC when the absolute lowest latency is required.

\vspace{-2ex}
\subsection{Programmable Offloading Engine}
\label{evaluation_smart_one_side}
\vspace{-1ex}

In this section, we evaluate the performance advantage of two programmable offloading functions. First is \textit{linked list traversal}; we consider a short linked list where each element contains an 8-bytes unique key and 64-bytes value. To locate a target key, the list is traversed starting from the head until a match is found, then returns the value pointed by the value pointer to the client. While RNIC performs the traversal on the client side, \sysname{} executes the traversal within the server-side network stack using lightweight intra-node DMA operations instead of expensive inter-node RDMA operations. Second is \textit{batched RDMA READ}. As the baseline, RNIC sends a series of 64-bytes RDMA READ packets. In our design, the client‑side network stack aggregates multiple target addresses into one consolidated request, after which the server‑side network stack issues concurrent DMA transfers to fetch all requested values in parallel and returns the aggregated data in a single response.

\begin{figure}[]
    \subfloat[Linked list traversal]{
        \label{fig_e_one_sided_linked_list}
        \includegraphics[keepaspectratio=true, width=0.50\linewidth]{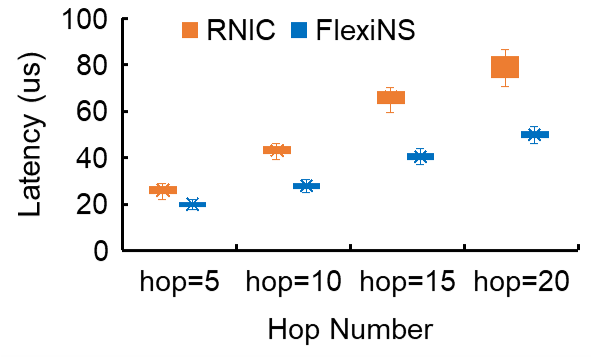}
    }
    \subfloat[Batched RDMA READ]{
        \label{fig_e_one_sided_batch_read}
        \includegraphics[keepaspectratio=true, width=0.50\linewidth]{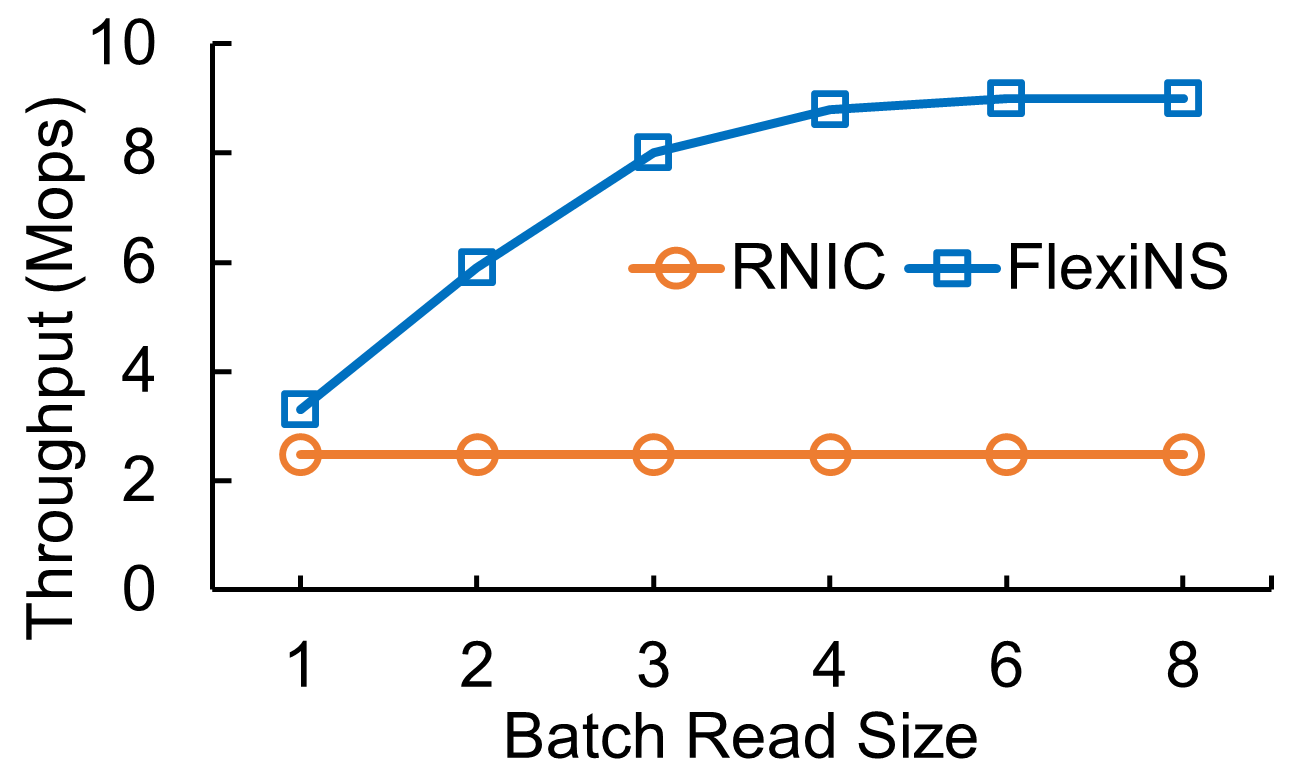}
    }
    \vspace{-1ex}
    \caption{Programmable offloading functions performance.}
    \label{fig_e_one_sided}
    \vspace{-3ex}
\end{figure}

Figure~\ref{fig_e_one_sided_linked_list} shows the traversal latency for varying hop counts, \sysname{} achieves 1.7$\times$ reduction in latency compared to RNIC, owing to the lightweight intra-node DMA operations, which incur lower latency than inter-node RDMA operations. \sysname{} also exhibits more stable latency performance, primarily because intra-node DMA operations are less susceptible to network fluctuations. Figure~\ref{fig_e_one_sided_batch_read} shows the throughput achieved by Batched RDMA READ on a single connection, \sysname{} achieved 3.5$\times$ higher throughput than RNIC, this is mainly because RDMA READ has limited outstanding number and simply increasing READ operations number doesn't increase throughput, but \sysname{} can perform READ concurrently with the powerful DMA engine, thereby enabling significantly enhanced throughput.

\vspace{-2ex}
\subsection{End-to-end Application Performance}
\label{evaluation_e2e}
\vspace{-1ex}


\noindent\textbf{Disaggregate block storage.} 
The compute‐storage separation architecture has become the de facto paradigm in modern data centers. In this design, compute servers and storage servers are organized into separate clusters, allowing each to be independently designed and optimized for specific workloads. The storage agent (SA), embedded within compute servers, converts storage operations into network transactions. Alibaba Solar~\cite{solar} is the transport protocol used in SA and has been widely deployed. Leveraging the high flexibility offered by \sysname{}, we implemented the Solar protocol in fewer than 2,000 lines of C++ code. As the baseline, we executed the protocol on dedicated CPU cores, similar to Snap. Additionally, we employ CRC NIC offload and Intel Data Streaming Accelerator~(DSA)~\cite{dsa} engine to further improve CPU performance. Similar to Luna~\cite{luna} and Solar~\cite{solar}, we select widely used 4KB READ request IOPS as metrics and limit the Solar-CPU network stack to 8 dedicated CPU cores. Each client is assigned to a dedicated core and set TX depth to 32.

\begin{figure}
	\centering
	{\includegraphics[keepaspectratio=true, width=0.9\linewidth]{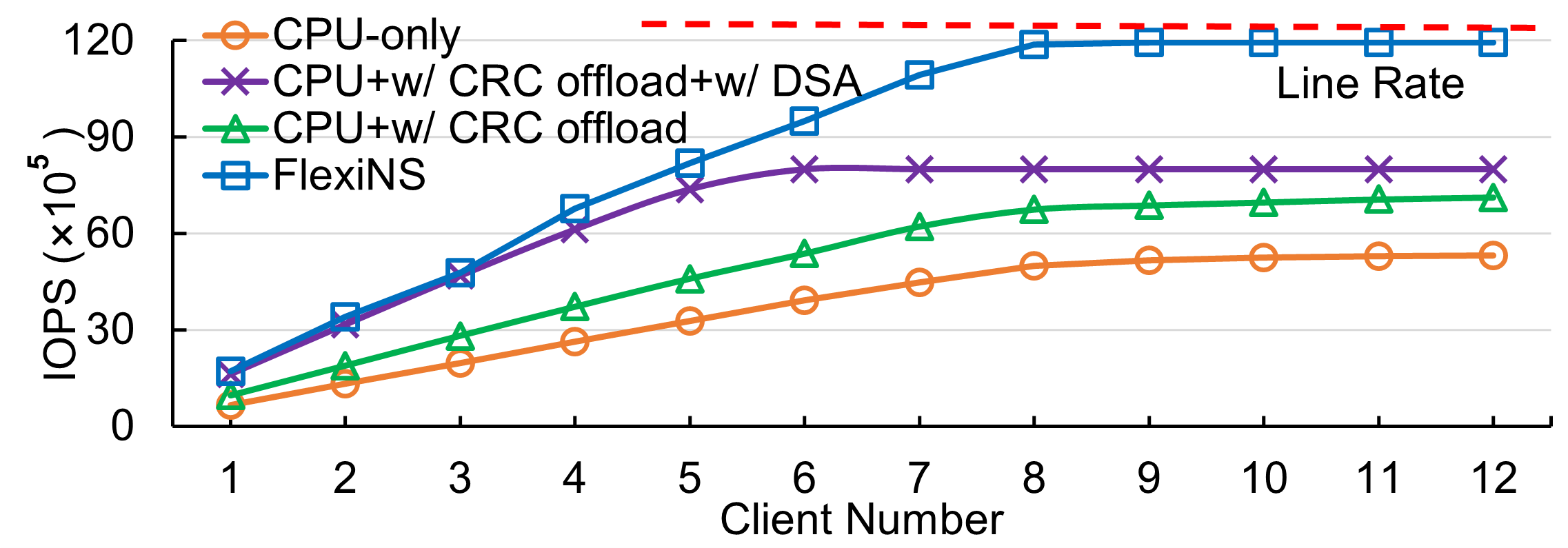}
        \vspace{-2ex}
	\caption{Performance comparison of four approaches when running the block storage application with Solar protocol.}
	\label{fig_e_e2e_solar}}
    \vspace{-4ex}
\end{figure}

Figure~\ref{fig_e_e2e_solar} illustrates the IOPS achieved by various approaches. We have two observations. First, \sysname{} reaches line rate with more than 8 clients, delivering 2.5$\times$ (2.2$\times$) higher IOPS compared to the "CPU-only" baseline for 1 client (12 clients). This is because \sysname{} offloads intensive CRC calculation to NIC hardware and massive memcpy to DMA engine, leaving only lightweight in-cache protocol processing for the Arm core. Second, even utilizing CRC offload and DSA engine, \sysname{} still achieves 1.5$\times$ higher IOPS than "Solar-CPU". Benefit from header-only offloading TX and unlimited-working-set in-cache processing RX, \sysname{} can easily operate at line rate, while "Solar-CPU" is constrained by the limitations of the CPU and DSA engine.

\noindent\textbf{KVCache Transfer.} 
Modern large language models (LLMs) are based on the Transformer architecture, and each inference request is logically divided into two stages: the prefill (P) stage and the decoding (D) stage. To meet stringent SLO requirements, the P/D disaggregation architecture is widely discussed and adopted~\cite{mooncake, deepseekv3, splitwise, distserve}. In this architecture, prefill nodes generate the KVCache and transfer it to the decoding nodes to continue decoding. Thus, efficient KVCache transfer from the prefill node pool to the decoding node pool is crucial. The Mooncake transfer engine~\cite{mooncake} provides a high-performance data transfer framework supporting RDMA/TCP protocols and Nvidia GPUDirect. However, Mooncake only chooses limited QP connections for each transfer task, which induces QP hash collision and one of the physical ports is underutilized (similar to ECMP hash collision~\cite{ecmp_hash,alibaba-hpn}). Therefore, we integrated the packet spraying mechanism~\cite{AWSSRD,mprdma} in \sysname{} that dynamically varies the source UDP port on each packet, substantially mitigating hash collisions and fully utilizing the bandwidth of both physical ports. We replace Mooncake RDMA with \sysname{} and compare it with the original Mooncake performance. Notably, \sysname{} also supports Nvidia GPUDirect, enabling the network stack to transfer payloads directly to/from pinned GPU memory. For our evaluation, we utilize 8 CPU cores to submit transfer tasks, bond two ports to one \texttt{mlx5\_bond\_0} port~\cite{link_aggr}, and set TX Depth to 1. 

\begin{figure}[]
    \subfloat[CPU mem to CPU mem]{
        \label{fig_e_e2e_kvtransfer_cpu2cpu}
        \includegraphics[keepaspectratio=true, width=0.50\linewidth]{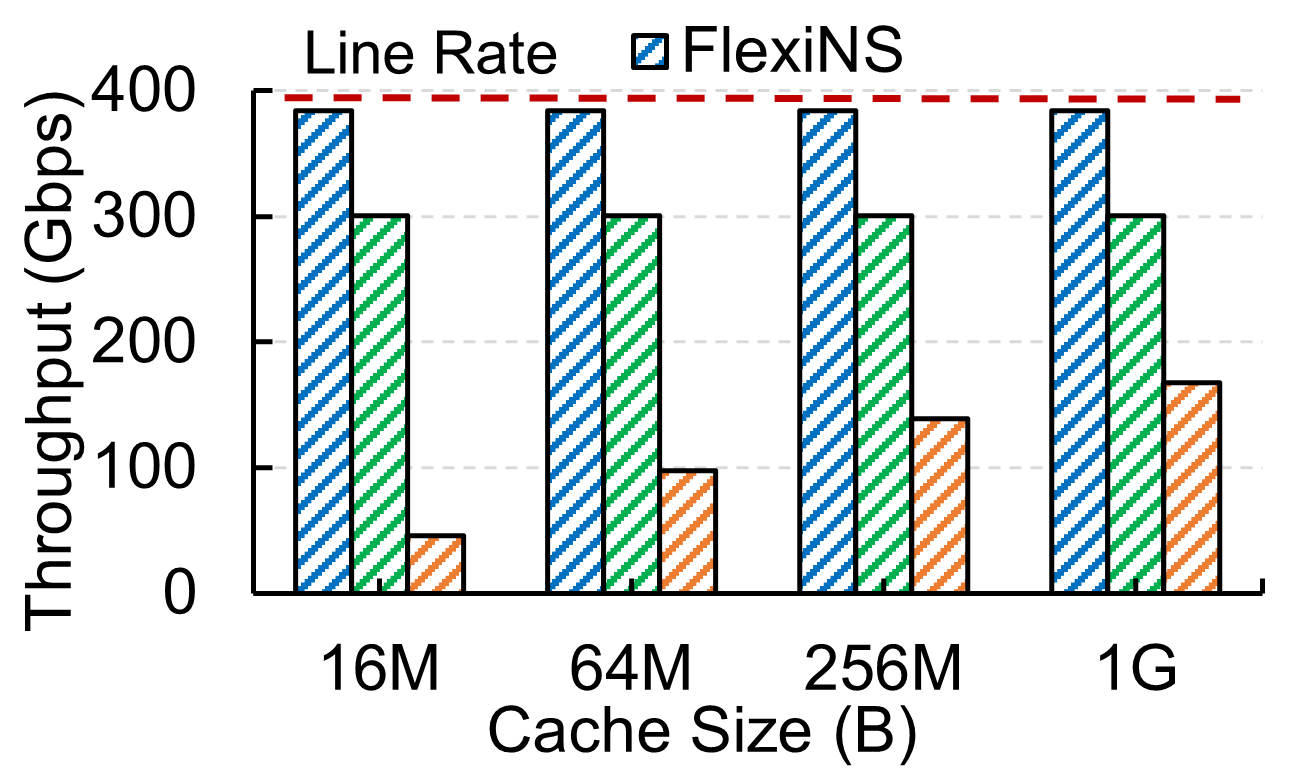}
    }
    \subfloat[CPU mem to GPU mem]{
        \label{fig_e_e2e_kvtransfer_gpu2gpu}
        \includegraphics[keepaspectratio=true, width=0.50\linewidth]{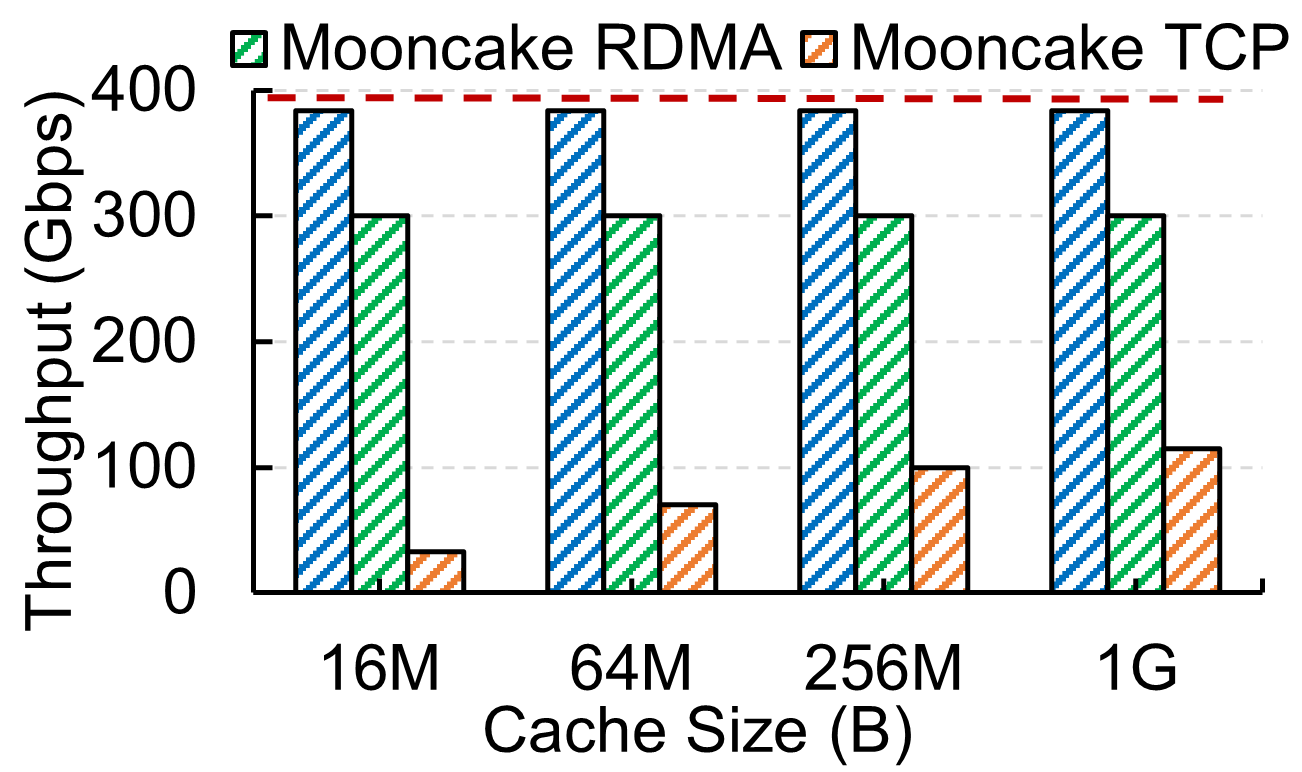}
    }
    \vspace{-2ex}
    \caption{Throughput comparison of KVCache transfer.}
    \label{fig_e_e2e}
    \vspace{-4ex}
\end{figure}

Figure~\ref{fig_e_e2e_kvtransfer_cpu2cpu} shows the transfer latency for various KVCache sizes between CPU memories, while Figure~\ref{fig_e_e2e_kvtransfer_gpu2gpu} shows the latency for transfers from CPU memory to pinned GPU memory. Leveraging header-only offloading TX and unlimited-working-set in-cache processing RX, \sysname{} achieves line rate performance for KVCache sizes exceeding 16MB, exhibits $3.2\times$ lower transfer latency compared to Mooncake TCP. Remarkably,  \sysname{} also outperforms Mooncake RDMA by $1.3\times$, attributed to the flexibility of \sysname{} can easily adopt the packet spraying mechanism to fully utilize the bandwidth of both ports. 

\vspace{-2ex}
\section{Related Work}
\vspace{-2ex}

\noindent\textbf{Network Stack as a Service.} 
IsoStack~\cite{isostack}, ZygOS~\cite{zygos}, TAS~\cite{tas}, NetKernel~\cite{NetKernel}, FreeFlow~\cite{freeflow}, and RoUD~\cite{roud} use dedicated CPU cores to construct an efficient and low latency network stack. Snap~\cite{Snap} is an industry framework proposed by Google and provided in a general-purpose, multi-tenant cloud environment. Shenango~\cite{shenango} designs a busy looping IOKernel and uses an entire core to thread scheduling and packet steering functions. SRM~\cite{qp_sharing_1} and KRCORE~\cite{KRCore} share QP connections in kernel mode and user mode, respectively, and perform as a network stack. However, they all introduce the extra memcpy and cause host overhead, memory bandwidth pressure, and application interference. In contrast, \sysname{} offloads the entire network stack to off-path SmartNIC, minimizing the host overhead and interference.

\noindent\textbf{Hardware-offloaded Network Stack.}
Extensive studies have offloaded the entire network stack to hardware and customized RDMA transport. From the industry, Google 1RMA~\cite{1RMA} and Falcon~\cite{Falcon}, AWS SRD~\cite{AWSSRD}, Meta~\cite{meta-rdma} and Microsoft Hyperscale ~\cite{azure-1,azure-2}, Alibaba Solar~\cite{solar} and HPN~\cite{alibaba-hpn}, are driving RDMA transport customizations at scale. This is supplemented by academic contributions like SRNIC~\cite{SRNIC} and StaR~\cite{StaR} focus on scalability, SmartDS~\cite{smartds} and RPCNIC~\cite{RPCAcc} focus on message split. All of them can achieve high throughput but offer low programmability and low flexibility. Instead, \sysname{} maintains line-rate and high flexibility through the novel designs of data and control paths.

\noindent\textbf{Offloading to SmartNIC.}
Offloading host workloads to SmartNICs has recently attracted significant attention in both academia and industry. Many prior works~\cite{tcp_offload_bf2,ipads_bf2,os2g} offload host tasks to off-path SmartNICs. IO-TCP~\cite{tcp_offload_bf2} offloads disk I/O and TCP packet transfer to SmartNIC and reduces the burden on the CPU for online content delivery. Xenic~\cite{xenic} offloads distributed transactions to SmartNIC. These works leverage SmartNIC to alleviate host CPU pressure but do not provide a comprehensive study on the offload network stack.


\vspace{-2ex}
\section{Conclusioin}
\vspace{-2ex}

This paper presents \sysname{}, a line-rate and flexible network stack with off-path SmartNIC. To tackle the ramifications of introducing SmartNIC, \sysname{} introduces a header-only offloading TX path, an unlimited-working-set in-cache processing RX path, a DMA-only notification pipe, and a programmable offloading engine. \sysname{} is immediately deployable, which maintains compatibility with IBV verbs and leverages off-the-shelf SmartNICs. The experimental results show that \sysname{} achieves 2.2$\times$ higher IOPS in block storage and 1.3 $\times$ higher throughput in KVCache transfer.

\bibliographystyle{plain}
\bibliography{content/ref}

\newpage
\appendix

\section{Appendix}

\subsection{One-sided Operation Support}
One-sided operations do not involve any user application logic on the destination, which has been widely used in lots of RDMA network systems~\cite{1RMA, farm,hybird-rdma-wei,one-side-disaggregated,rdma-design-guide, storm}. 

\sysname{} natively supports RDMA one-sided operations: When the server network stack receives a READ operation, it constructs the response header and utilizes the header-only offloading TX path ($\S$\ref{sec:tx_path}) to return the target payload to the client network stack, which then transfers the payload to the user application buffer. When the server network stack receives a WRITE operation, it directly transfers the payload to the target user application buffer and sends an ACK to the client network stack. Notably, both operations eliminate host CPU involvement.

\subsection{Implementation Details}
\label{implement_details}
\noindent\textbf{RDMA IBV Verbs compatibility.}
IBV Verbs~\cite{rdma-core} is the de facto standard in RDMA programming. Unlike many prior works that design custom API frameworks, we register the entire system as an RDMA device within the kernel module and implement commonly used IBV Verbs interfaces in both kernel and user-space libraries, including \textbf{13 control path verbs} and \textbf{4 data path verbs}. With minimal code modifications, developer applications can leverage the flexibility and high throughput provided by \sysname{}.

\noindent\textbf{Core partition.} 
To achieve higher process efficiency, the Arm cores are divided into data and control cores. Data cores execute the common‑case network stack operations, managing all flows TX/RX tasks with minimal overhead. Control processors execute control verbs, allocate essential resources (e.g., shadow memory regions and notification pipes), and communicate with kernel modules. In addition, control cores maintain a retransmission timer for each flow that aggregates unacknowledged data from the data cores and a congestion controller that updates the transmission rate by customized CCA.

\noindent\textbf{DOCA-Free system.}
NVIDIA DOCA framework~\cite{doca} is a commonly used SDK that simplifies BF3 development. However, to reduce the complexity of deployment dependencies while improving DMA and cache operation performance, we abandoned the DOCA framework and directly utilized the NVIDIA DevX interface~\cite{devx} and IBV Verbs interface for the core IOKernel and the user-space libraries.

\noindent\textbf{Cache Invalidate operation.}
Although ARMv7 ISA already provides DCIMVAC operation~\cite{idio}, we don't directly use this mechanism. Instead, BF3 provides the specific NIC opcode to facilitate invalidating a set of cache lines~\cite{cache_invalid}, thus we construct the corresponding WQE and use MMIO to write WQE into the NIC UAR page, which introduces only minimal Arm core overhead. In addition, we engaged in private discussions with Mellanox engineers regarding combining the DMA opcode and the cache invalidation opcode into a single opcode, which further reduces the overhead on the Arm core and will be released near future.

\subsection{Programmable offloading engine example}
\label{codeing_example}

\begin{lstlisting}[caption={A coding example of programming with programmable offloading engine},captionpos=b]
#define BATCH_READ_OPCODE 0x1234

#define VALUE_SIZE 64

void register_batch_read() {
  /* create a RC qp */
  ibv_qp *qp = create_rc_qp();

  /* connect this qp with remote qp */
  connect_rc_qp(qp);

  /* exchange with host to get the target address and size */
  (host_addr, size) = exchange_data_with_host();

  void *host_buffer = register_dma_region(host_addr, size);

  register_opcode(BATCH_READ_OPCODE, qp, handle_batch_read);
}

void handle_batch_read(void *packet, int length, qp_context *context) {
  /* assume packet[0] contains total read number,
      and following the target read address */
  int read_number = packet[0];

  void *resp_buffer = alloc_resp(read_number * VALUE_SIZE);

  int dma_id[read_number];
  for (int i = 0; i < read_number; i++) {
    int addr_offset = packet[1 + i];
    /* submit DMA to read host buffer to resp_buffer */
    dma_id[i] = submit_dma(context, READ, context->host_addr + addr_offset,
                           resp_buffer + i * VALUE_SIZE, VALUE_SIZE);
  }

  for (int i = 0; i < read_number; i++) {
    /* wait all dma operations finish */
    wait_dma_finish(dma_id[i]);
  }

  submit_resp(context, resp_buffer, read_number * VALUE_SIZE);
}
\end{lstlisting}

\end{sloppypar}
\end{document}